\documentclass
[twocolumn,aps,prc,amsmath,showpacs,amssymb,floatfix]
{revtex4}
\usepackage{amssymb}

\usepackage{CJK}                     % chinese fonts
\usepackage[dvips]{graphicx}
\usepackage{bm}                      % bold math
\usepackage{mathptmx}                % math+times fonts
\usepackage{dcolumn}                 % Align table columns on decimal point
\usepackage{array}
\usepackage{graphicx}% Include figure files
\usepackage{dcolumn}% Align table columns on decimal point
\usepackage{bm}% bold math
\usepackage{ulem} %% for strike-through
\usepackage[usenames]{color}
\usepackage{epstopdf}
\usepackage{epsfig}
\usepackage{float}
\usepackage{subfigure}
\usepackage{multirow}
\usepackage{slashed}
\usepackage{diagbox}
\usepackage{cancel}
\usepackage{graphics}
\usepackage{longtable,pdflscape}
\newcommand{\nc}{\newcommand}       % new command
\nc{\vc}[1] {\mbox{\boldmath $#1$}} % boldmath(vector)
\nc{\del}       {\partial}              % bra state
\nc{\bra}       {\langle}               % bra state
\nc{\ket}       {\rangle}               % ket state
\nc{\bras}[1]   {\langle #1|}           % bra state
\nc{\kets}[1]   {|#1\rangle}            % ket state
\nc{\mapleft}[1]{           % something under arrow
 \smash{\mathop{\,          %
  \hbox to 1.5cm{\rightarrowfill}\, }\limits_{#1}}}
\nc{\beq}     {\begin{eqnarray}} \nc{\eeq}    {\end{eqnarray}}
\nc{\nn}      {\\\nonumber} \nc{\vs}      {\vspace{-0.275cm}}
\nc{\fra}    {\frac{1}{2}}
\nc{\mb}        {\mathbf}
\usepackage{color}

\begin{document}

\title{$\Delta$ (1232) effects in density-dependent relativistic Hartree-Fock theory and neutron stars}
\author{Zhen-Yu Zhu$^{1}$, Ang Li$^{1}$\footnote{liang@xmu.edu.cn}, Jin-Niu Hu$^{2}$\footnote{hujinniu@nankai.edu.cn}, Hiroyuki Sagawa$^{3,4}$}
\affiliation{
$^1$Department of Astronomy, Xiamen University, Xiamen 361005, China\\
$^2$School of Physics, Nankai University, Tianjin 300071, China\\
$^3$RIKEN Nishina Center, RIKEN, Wako 351-0198, Japan\\
$^4$Center for Mathematics and Physics, University of Aizu, Aizu-Wakamatsu, Fukushima 965-8560, Japan\\
}
\date{\today}

\begin{abstract}
The density-dependent relativistic Hartree-Fock (DDRHF) theory is extended to include $\Delta$-isobars for the study of dense nuclear matter and neutron stars. To this end, we solve the Rarita-Schwinger equation for spin-3/2 particle. Both the direct and exchange terms of the $\Delta$-isobars' self-energies are evaluated in details. In comparison with the relativistic mean field theory (Hartree approximation), a weaker parameter dependence is found for DDRHF. An early appearance of $\Delta$-isobars is recognized at $\rho_B\sim0.28$fm$^{-3}$, comparable with that of hyperons. Also, we find that the $\Delta$-isobars' softening of the equation of state is mainly due to the reduced Fock contributions from the coupling of the isoscalar mesons, while the pion contributions are negligibly small. We finally conclude that with typical parameter sets, neutron stars with $\Delta$-isobars in their interiors could be as heavy as the two massive pulsars whose masses are precisely measured, with slightly smaller radii than normal neutron stars.
\end{abstract}

% 21.60.Jz  % Nuclear Density Functional Theory and extensions (includes Hartree-Fock and random-phase approximations)
% 21.30.Fe   % Forces in hadronic systems and effective interactions
% 26.60.-c,  % Nuclear matter aspects of neutron stars
% 26.60.Kp,  % Equations of state of neutron star matter
% 21.65.Mn,  % Equations of state of nuclear matter
% 26.50.+x,  % Nuclear physics aspects of supernovae and other explosive environments
% 26.60.Dd,  % Neutron star core
% 26.60.Gj,  % Neutron star crust
% 97.60.Jd,  % Neutron stars
% 21.65.Mn,  % Equations of state of nuclear matter
% 21.65.+f,  % Nuclear matter
% 21.65.Cd,  % Asymmetric matter, neutron matter
% 24.10.Cn,  % Many-body theory
% 13.75.Cs,  % Nucleon-nucleon interactions

%\keywords{Delta isobar \sep Relativistic Hartree-Fock model \sep Neutron star}
%Use showkeys class option if keyword
%display desired

\maketitle

\section{Introduction}

Recently the early appearance of $\Delta$-isobars in dense nuclear matter has intrigued many illuminating studies for the interest of both neutron stars (NSs)~\cite{drago14prc,schramm10apj,drago14prd,guo03prc,li15prc,delta} and heavy ion collision~\cite{hic}. In particular, very compact stellar configurations~\cite{drago14prc,schramm10apj} are reached due to the introduction of $\Delta$-isobars as required by some current radius measurements~\cite{smallradius1,smallradius2,smallradius3}. However, there is the $\Delta$ puzzle~\cite{drago14prc} similar with the hyperon puzzle~\cite{hpuzzle,Li11y, rijken} that the corresponding maximum NSs' masses could not fulfill the constraint from the recent two precisely-measured 2-solar-mass pulsars PSR J1614-2230~\cite{2solar10,2solar11} and PSR J0348+0432~\cite{2solar2}.

Previous studies have suggested the two-family scenario of compact stars to resolve this problem~\cite{drago14prd}, we here go beyond the previously-employed relativistic mean field (RMF) theory to include the contributions due to the exchange (Fock) terms and the pseudo-vector $\pi$-meson couplings which are effective only through exchange
terms. Therefore, we employ the density-dependent relativistic Hartree-Fock (DDRHF) theory~\cite{bouyssy87prc} to study this problem.

As one of the most advanced nuclear many-body model based on the covariant density functional theory, DDRHF presents a quantitative description of nuclear phenomena~\cite{long12prc,sun08prc,long06plb,long08el,ddrhf1,ddrhf2,ddrhf3} with a similar accuracy as RMF. In DDRHF, the Lorentz covariant structure is kept in full rigor, which guarantees all well-conserved relativistic symmetries. Previously, it has been demonstrated
that the isoscalar Fock terms are essential for the prediction of NS properties~\cite{long12prc,sun08prc}. Also, we expect that the density-dependence~\cite{long06plb,rmf99npa} introduced in the present study for all meson couplings would make difference on the $\Delta$-matter study.

Besides the $\Delta$-puzzle problem in NSs, we are also interested in how the DDRHF results depend on the uncertain $\Delta$-meson coupling constants, since there are hardly previous studies except with RMF. Presently, there is no constraint on $\Delta$-$\rho$ coupling. Several constraints exist for $\sigma, \omega$ couplings from the quark model~\cite{1prd74,1npa79,plb81x}, finite-density QCD sum-rule methods~\cite{qcd95}, quantum hadrodynamics~\cite{prc12x,npa89x}, and laboratory experiments~\cite{prc12xx,prc97x,plb05x}. Based on the quark counting argument~\cite{1prd74,1npa79}, there are the universal couplings between nucleons, $\Delta$-isobars with mesons, namely, $x_{\lambda} = g_{\lambda\Delta}/g_{\lambda N} = 1$ ($\lambda$ labels meson). A theoretical analysis~\cite{plb81x} of M1 giant resonance and Gamow-Teller transitions in nuclei found $25-40\%$ reductions of the transition strength due to the couplings to $\Delta$ isobars,
%$N\Delta$ strength over $NN$ strength based on the quark model,
while recent experiments~\cite{prc12xx,prc97x,plb05x} bring down the quenching value to be at most $10-15\%$ due to the coupling to $\Delta$-isobars. This means that the  $\Delta$ couplings of isoscalar mesons ($\sigma, \omega$) should be  weaker than those of quark model prediction, namely $x_{\sigma},  x_{\omega}\leq1$ are suggested by these experiments. Also, at the saturation density $\rho_0$, a possible smaller $\Delta$-isobars' vector self-energy was indicated~\cite{qcd95} than the corresponding value for the nucleon [i.e., $x_{\omega}(\rho_0)<\sim1$], while the $\Delta$-isobars' scalar self-energy was difficult to predict. In the analysis of Ref.~\cite{npa89x} in Hartree approximation, the difference between $x_{\sigma}$ and $x_{\omega}$ was found to be $x_{\omega}=x_{\sigma}-0.2$. A recent study, however, concluded that the cross sections involved are not sensitive to $x_{\omega}$ and $x_{\sigma}$ either~\cite{prc12x}. In the present study, for the purpose of comparing with previous RMF studies~\cite{li15prc}, we scale the density-dependence of $\Delta$-meson coupling in a reasonable range according to previous constraints for $\sigma, \omega, \rho$ mesons. The $\pi$ meson coupling is fixed to be $x_\pi=1.0$ hereafter.

We provide the DDRHF formula extended to include $\Delta$-isobars in Sect.~II. Our results and discussions will be given in Sect.~III, before drawing conclusions in Sect.~IV.

%-------------------------------------------------------------------------------
\section{DDRHF model with $\Delta$-isobars}

Here we extend the DDRHF calculation to $\Delta$-matter. The model system consists of nucleons $\psi_N$, $\Delta$-isobars ($\Delta^{-}, \Delta^{0},\Delta^{+},\Delta^{++}$) which are treated as a Rarita-Schwinger particle $\psi_{\Delta}^\alpha$, and the effective fields of two isoscalar mesons ($\sigma$ and $\omega$) as well as two isovector mesons ($\pi$ and $\rho$). The effective Lagrangian density can be written as three parts:
\beq
\mathcal{L}=\mathcal{L}_0 + \mathcal{L}_{IN} + \mathcal{L}_{I\Delta},
\eeq
where the $\mathcal{L}_0$ denotes the free Lagrangian density of baryons and mesons, $\mathcal{L}_{IN}$ denotes the interaction Lagrangian density between mesons and nucleons, and $\mathcal{L}_{I\Delta}$ denotes the interaction Lagrangian between mesons and $\Delta$-isobars. More physics details can be found in Ref.~\cite{bouyssy87prc}, and we write explicitly here only the $\mathcal{L}_{I\Delta}$ term which is newly introduced in the present work:
\beq
\mathcal{L}_{I\Delta} &= & -g_{\sigma\Delta} \bar{\psi}_{\Delta\alpha} \sigma\psi_\Delta^\alpha \nonumber \\
% sigma
&&- g_{\omega_\Delta} \bar{\psi}_{\Delta\alpha} \gamma_\mu \omega^\mu \psi_\Delta^\alpha                       % omega V
+ \frac{f_{\omega_\Delta}}{2M_\Delta}\bar{\psi}_{\Delta\alpha}
\sigma_{\mu\nu}\partial^\nu \omega^\mu \psi_\Delta^\alpha \nonumber \\                                                 % omega T
&&- g_{\rho_\Delta} \bar{\psi}_{\Delta\alpha} \gamma_\mu \bm{\rho}^\mu \cdot
\bm{\tau_\Delta}\psi_\Delta^\alpha                                                            % rho V
 + \frac{f_{\rho_\Delta}}{2M_\Delta}\bar{\psi}_{\Delta\alpha} \sigma_{\mu\nu}\partial^\nu \bm{\rho}^\mu\cdot \bm{\tau_\Delta}\psi_\Delta^\alpha \nonumber \\                                                                         % rho T
&&- \frac{f_{\pi_\Delta}}{m_\pi}\bar{\psi}_{\Delta\alpha} \gamma_5 \gamma_\mu \partial^\mu \bm{\pi}\cdot \bm{\tau_\Delta}\psi_\Delta^\alpha,                                                                        % pi
\eeq
where the $\Delta$ isospin operator $\bm{\tau_\Delta}$ has the form of a $4\times 4$ matrix. $M_\Delta$ denotes the bare $\Delta$-isobar mass. We follow Ref.~\cite{bouyssy87prc} and neglect the small tensor coupling of $\omega$ meson for both nucleons and $\Delta$-isobars in our calculations.

From the Lagrangian in Eq.~(1), following the standard procedure in Ref.~\cite{bouyssy87prc}, one can derive the Hamiltonian as:
\beq
&& H = \int_{t=0}\bar{\psi}_N(-i\bm{\gamma}\cdot \nabla + M_N)\psi_N d^3x   \nonumber \\
&& + \int_{t=0} \bar{\psi}_{\Delta \alpha}(-i\bm{\gamma}\cdot \nabla + M_\Delta)
\psi_\Delta^\alpha d^3x \nonumber \\                                                                         % kinetic
&& +
\frac{1}{2}\sum_{\lambda}\int_{t=0}\mathop{d^3xd^4x'}\bar{\psi}_a(x)\bar{\psi}_b\mathop(x') \Gamma_{\lambda}^{ab}D_{\lambda}\mathop(x-x')\psi_b\mathop(x')\psi_a(x),\nonumber\\                                 % potential
\eeq
where the index $\lambda$ represents each meson ($\lambda = \sigma, \omega, \pi, \rho$), and $a,b$ represent nucleon or $\Delta$-isobar. $\Gamma_{\lambda}^{ab}$ are the vertices between mesons and baryons as listed in details in Eqs.~(43-48) of Appendix A. $D_\lambda(\mathop{x-x'})$ are the propagator of each meson, respectively.

In the mean field approximation, the field operator of nucleon should satisfy the Dirac equation:
\beq
(-i\gamma^\mu \partial_\mu + M_N + \Sigma_N)\psi(x)=0,
\eeq
while the field operator of $\Delta$-isobars must be described by the Rarita-Schwinger equations as spin $3/2$ particle,
 \beq
(-i\gamma^\mu \partial_\mu + M_\Delta + \Sigma_\Delta)\psi(x)^\alpha=0,~~~~\gamma_\alpha\psi^\alpha(x)=0,
\eeq
with the corresponding self-energies $\Sigma_N, \Sigma_\Delta$ to be determined self-consistently. $M_N$ is the bare nucleon mass.

The field operators of nucleons and $\Delta$-isobars could be expanded in the second quantization forms:
\beq
\psi(\bm{x},t)&=& \sum_\kappa\left[f_\kappa(\bm{x})e^{-iE_\kappa t}b_\kappa^N
+ g_\kappa(\bm{x})e^{i\mathop{E'_\kappa} t}d_\kappa^{N\dagger}\right],   \\
\psi^\dagger(\bm{x},t) &=&\sum_\kappa\left[f_\kappa^\dagger(\bm{x})e^{iE_\kappa t}b_\kappa^{N\dagger}
+ g_\kappa^\dagger(\bm{x})e^{-i\mathop{E'_\kappa} t}d_\kappa^N\right],   \\                                  % nucleon
\psi^\alpha(\bm{x},t) &=& \sum_\kappa\left[f_\kappa^\alpha(\bm{x})e^{-iE_\kappa t}b_\kappa^\Delta
+ g_\kappa^\alpha(\bm{x})e^{i\mathop{E'_\kappa} t}d_\kappa^{\Delta\dagger}\right],\\
\psi_{\alpha}^\dagger(\bm{x},t)&=& \sum_\kappa\left[f_{\kappa\alpha}^\dagger(\bm{x})e^{iE_\alpha t}b_\kappa^{\Delta\dagger}
+ g_{\kappa\alpha}^\dagger(\bm{x})e^{-i\mathop{E'_\kappa} t}d_\kappa^\Delta \right].
\eeq
Here $b_\kappa^{(N/\Delta)}$ and $b_\kappa^{(N/\Delta)\dagger}$ are the annihilation and creation operators for nucleons$/\Delta$-isobars, while $d_\kappa^{(N/\Delta)}$ and $d_\kappa^{(N/\Delta)\dagger}$ are the annihilation and creation operators for the corresponding anti-particles, respectively. The no-sea approximation is adopted in this work. Then the contributions of antiparticles are neglected, and only the particle terms of the baryon field operators are reserved.

For nuclear matter, the Dirac spinor $f_\kappa(\bm{x})$ or Rarita-Schwinger spinor $f_\kappa^\alpha(\bm{x})$ could be written as:
\beq
f_\kappa(\bm{x}) &=& u(\bm{p},\kappa)e^{i\bm{p}\cdot\bm{x}},\\
f_\kappa^\alpha(\bm{x}) &= &u^\alpha(\bm{p},\kappa)e^{i\bm{p}\cdot\bm{x}},
\eeq
where the $u(\bm{p},\kappa)$ and $u^\alpha(\bm{p},\kappa)$ are the solutions of the following equations, respectively:
\beq
(\bm{\gamma}\cdot \bm{p} + M_N + \Sigma_N)u(\bm{p},\kappa) &=& \gamma^0 Eu(\bm{p},\kappa), \label{diracf} \\
(\bm{\gamma}\cdot \bm{p} + M_\Delta + \Sigma_\Delta)u^\alpha(\bm{p},\kappa) &=& \gamma^0 Eu^\alpha(\bm{p},\kappa).
\eeq
Here $E$ is the single particle energy in nuclear matter. $\kappa$ contains both spin $s$ and isospin $\tau$ components. From now on, we omit for simplicity the index $\tau$.

In consistent with the rotational invariance of the infinite nuclear matter, the baryon self-energy $\Sigma_i~(i=n,~p, ~\Delta^-,~\Delta^0,~\Delta^{+},~\Delta^{++})$, produced by
the meson exchanges, can be written quite generally as
\beq
\Sigma_{i}(p) = \Sigma_S^{i}(p) + \gamma_0\Sigma_0^{i}(p) + \bm{\gamma}\cdot\bm{\hat{p}}\Sigma_V^{i}(p)
\eeq
where $\Sigma_S, ~\Sigma_0, ~\Sigma_V$ are respectively the scalar, timelike and spacelike-vector components of the self-energy.

Then Eq.~(12) is simplified as
\beq
(\bm{\gamma}\cdot \bm{p}^\ast + M^\ast)u(\bm{p},s) = \gamma_0 E^\ast u(\bm{p},s),
\eeq
where the starred quantities are defined by
\beq
\bm{p}^\ast(p) &=& \bm{p} + \bm{\hat{p}}\Sigma_V(p),\\
 M^\ast(p) &=& M + \Sigma_S(p),\\
 E^\ast(p) &=& E(p) - \Sigma_0(p).
\eeq
$M^\ast$ is the usual scalar effective mass of the baryon, and $E^{\ast2} = \bm{p}^{\ast2}  + M^{\ast2} $.

Eq.~(15) has the similar form with the Dirac equation of a free particle. The positive energy solution is
\beq
u(\bm{p},s) = \left(\frac{E^\ast + M^\ast}{2E^\ast}\right)^{1/2} \begin{bmatrix}
1 \\ \cfrac{\bm{\sigma}\cdot\bm{p}^\ast}{E^\ast + M^\ast}
\end{bmatrix}\chi_s,
\eeq
where $\chi_s$ is the wave function of spin. The normalization condition of Dirac spinor is chosen as
\beq
u^\dagger(\bm{p},s)u(\bm{p},s)=1.
\eeq

While for the $\Delta$-isobars, $u^\alpha(\bm{p},s_\Delta)$ can be obtained from the angular momentum coupling of spin 1 and spin $1/2$ as follows:
\beq\label{raritaspin}
u^\alpha(\bm{p},s_\Delta) =\sum_{s_1,s_2}(1s_1\frac{1}{2}s_2|\frac{3}{2}s_\Delta) \epsilon^\alpha(\bm{p},s_1)u(\bm{p},s_2),
\eeq
where $(1s_1\frac{1}{2}s_2|\frac{3}{2}s_\Delta)$ is the Clebsch-Gordan coefficient and $\epsilon^\alpha(\bm{p},s_1)$ is the unit vector of spin 1 particle. Introducing the detailed values of Clebsch-Gordan coefficients in Eq.~(\ref{raritaspin}), one can write explicitly the wave function of $\Delta$-isobars for all spin components:
\beq
 u^\alpha(\bm{p},\pm \frac{3}{2}) &=& \epsilon^\alpha(\bm{p},\pm 1)u(\bm{p},\pm \frac{1}{2}), \\
 u^\alpha(\bm{p},\pm \frac{1}{2}) &=& \sqrt{\frac{2}{3}}\epsilon^\alpha(\bm{p},0)u(\bm{p},\pm \frac{1}{2})
 \nonumber \\
 && + \sqrt{\frac{1}{3}}\epsilon^\alpha(\bm{p},\pm 1)u(\bm{p},\mp \frac{1}{2}).\nonumber \\
\eeq
The unit vector $\epsilon^\alpha(\bm{p},s_1)$ are determined by
\beq
\gamma_\alpha u^\alpha(\bm{p},s_\Delta) = 0.
\eeq
The normalization condition of the Rarita-Schwinger spinor satisfies
\beq
u_\alpha^{\dag}(\bm{p},s_\Delta)u^{\alpha}(\bm{p},s_\Delta)=-1.
\eeq

\begin{figure}
\vspace{0.3cm}
{\centering
\resizebox*{0.48\textwidth}{0.3\textheight}
{\includegraphics{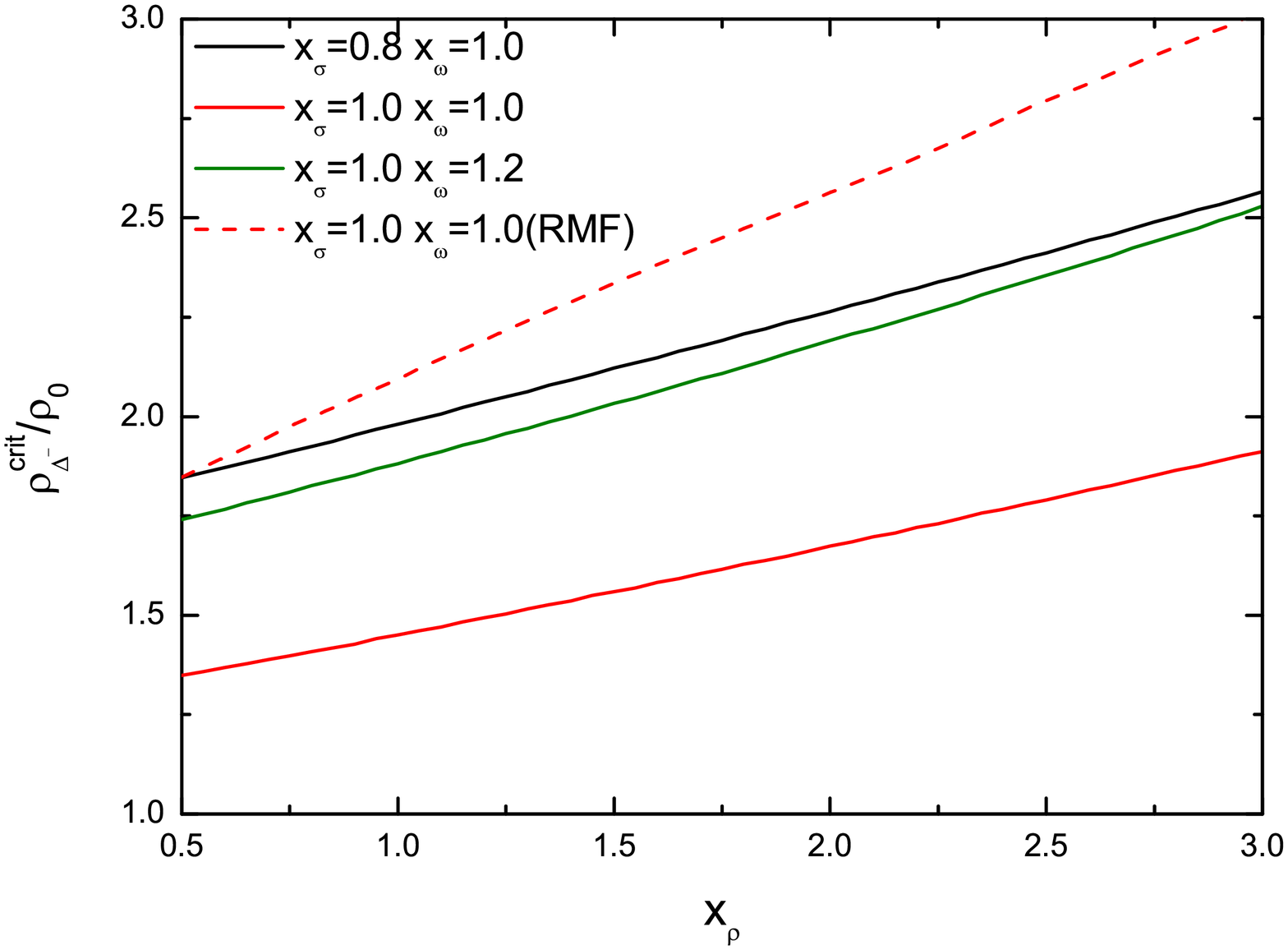}}
\par}
\caption{\small(Color online) $\Delta$ critical density $\rho^{\rm crit}_{\Delta}$ as a function of $x_\rho$, for $(x_\sigma, x_\omega) = (0.8, 1.0), (1.0, 1.0), (1.0, 1.2)$. Previous RMF calculations~\cite{li15prc} with $(x_\sigma, x_\omega) = (1.0, 1.0)$ are also shown for comparison.}\label{fig1}
\end{figure}

Next we define some useful quantities as
\beq
\hat{P} \equiv \frac{|\bm{p}^\ast|}{E^\ast},~~~ \hat{M} \equiv \frac{M^\ast}{E^\ast}.
\eeq
They can also be expressed by $u(\bm{p},s)$ and $u^\alpha(\bm{p},s_\Delta)$:
\beq
 \hat{M}_N(\bm{p}) &=& \frac{1}{2}\sum_s \bar{u}(\bm{p},s) u(\bm{p},s),  \\
 \hat{P}_N(\bm{p}) &=& \frac{1}{2}\sum_s \bar{u}(\bm{p},s)\bm{\gamma}\cdot \hat{\bm{p}} u(\bm{p},s);\\
 \hat{M}_\Delta(\bm{p}) &=& -\frac{1}{4}\sum_{s_\Delta} \bar{u}_\alpha(\bm{p},s_\Delta) u^\alpha(\bm{p},s_\Delta),\\
 \hat{P}_\Delta(\bm{p}) &=& -\frac{1}{4}\sum_{s_\Delta} \bar{u}_\alpha(\bm{p},s_\Delta)\bm{\gamma}\cdot \hat{\bm{p}} u^\alpha(\bm{p},s_\Delta),
\eeq
where the coefficient $1/4$ in Eqs.~(29-30) is generated by spin $3/2$.

The energy density of nuclear matter will be obtained through calculating the Hamiltonian density with the Hartree-Fock wave function of baryons,
\beq
|HF\rangle = \prod b^\dagger(p,\kappa)|0\rangle,
\eeq
where $|0\rangle$ is the physical vacuum. With this wave function, the energy density of nuclear matter could be written as:
\beq
\varepsilon = \frac{1}{\Omega}\langle HF|H|HF\rangle = \langle T\rangle + \langle V\rangle,
\eeq
where $\Omega$ is the volume of the system. $\langle T\rangle$ and $\langle V\rangle$ are the kinetic part and potential part of the energy density, respectively. $\langle V\rangle$ can be further decomposed into direct (Hartree) term $\langle V_D\rangle$ and exchange (Fock) term $\langle V_E\rangle$ (See details in Appendix B for the calculations of the energy density).

\begin{figure}
\vspace{0.3cm}
{\centering
\resizebox*{0.48\textwidth}{0.3\textheight}
{\includegraphics{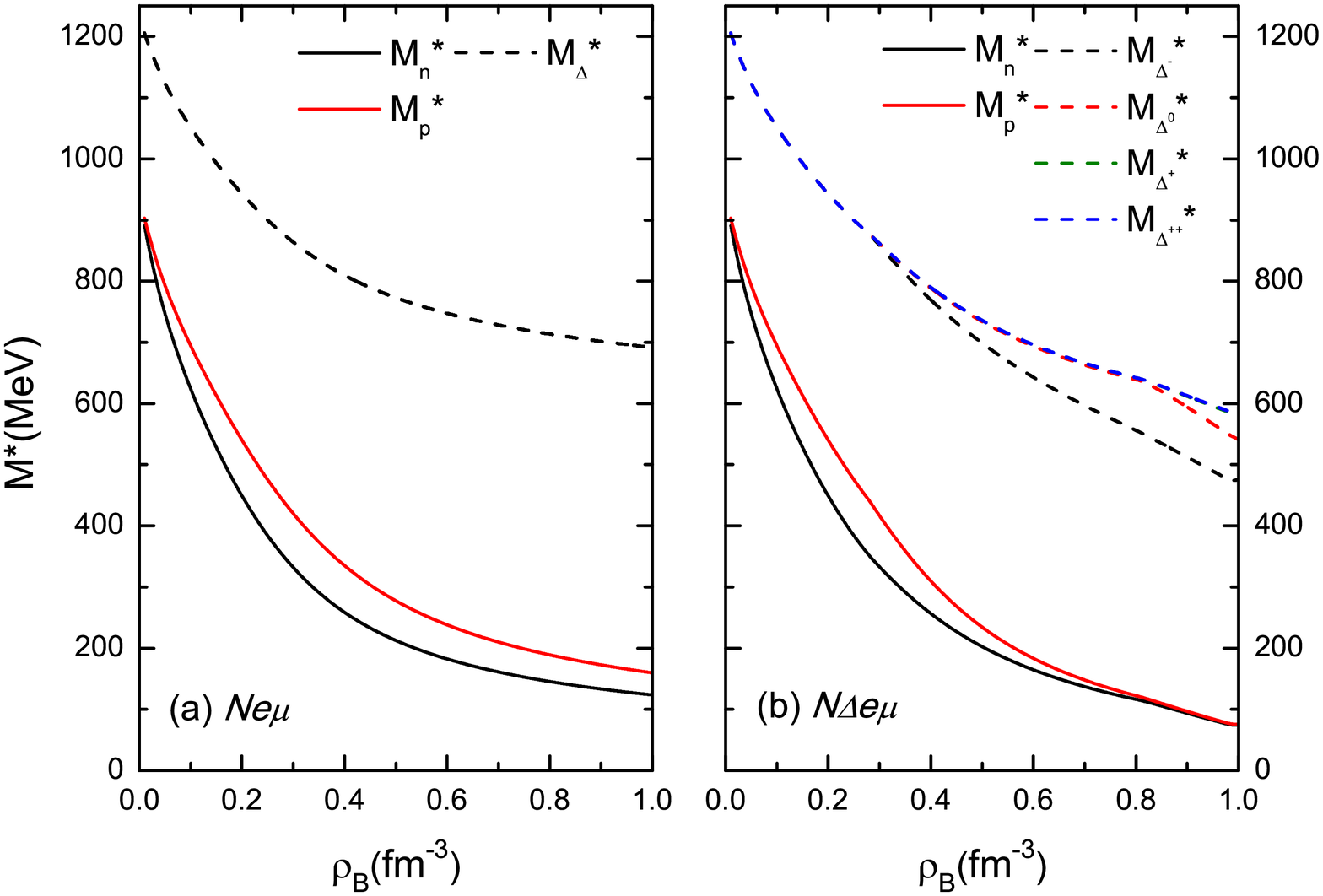}}
\par}
\caption{\small(Color online) Effective baryonic mass $M^*$ as a function of the baryon density $\rho_B$ for both $Ne\mu$ matter and $N\Delta e\mu$ matter. The calculations are done for one representative parameter set (PKO1) at fixed meson coupling of $(x_\sigma, x_\omega, x_{\rho}) = (0.8, 1.0, 0.5)$.}\label{fig2}
\end{figure}

The self-energies for nucleons/$\Delta$-isobars are obtained by differentiating $\langle V\rangle$ with respect to $u(\bm{p},s)$/$u^\alpha(\bm{p},s_\Delta)$. The results for all self-energy components are presented in Appendix D.

For neutron star matter, there are not only baryons but also leptons (electrons and muons). They are in equilibrium under weak processes, which leads to the following relations among the involved chemical potentials:
\beq
\mu_n=\mu_p+\mu_e, &~& \mu_\mu=\mu_e,\\
\mu_n=\mu_{\Delta^-}-\mu_e,&~& \mu_n=\mu_{\Delta^0},\\
\mu_n=\mu_{\Delta^+}+\mu_e,&~&
\mu_n=\mu_{\Delta^{++}}+2\mu_e.
\eeq
The chemical potentials are determined by the relativistic energy-momentum relation at Fermi momentum ($p=k_F$),
\beq
\mu_i&=&\Sigma^i_0(k_{F,i})+E^\ast_i(k_{F,i}),\\
\mu_l&=&\sqrt{k^2_{F,l}+m^2_l},
\eeq
where $i=n,~p, ~\Delta^-,~\Delta^0,~\Delta^{+},~\Delta^{++}$ and $l=e,\mu$.
Furthermore, the baryon number conservation and charge neutrality are imposed in neutron star matter as,
\beq
&&\rho_B=\rho_{Bn}+\rho_{Bp}+\rho_{B\Delta^-}+\rho_{B\Delta^0}+\rho_{B\Delta^+}+\rho_{B\Delta^{++}},\\
&&\rho_e+\rho_\mu=2\rho_{B\Delta^{++}}+\rho_{B\Delta^{+}}+\rho_{Bp}.
\eeq
The pressure is calculated following the thermodynamics relation as
\beq
P(\rho_B)=\rho^2_B\frac{\partial}{\partial\rho_B}\frac{\varepsilon}{\rho_B}=\sum_{j=i,l}\rho_{Bj}\mu_j-\varepsilon
+\rho^2_B\sum_{j=i,l}\mu_j\frac{dY_j}{d\rho_B}.
\eeq
Here $Y_j=\rho_j/\rho_B$ is the relative fraction of each particle species $j$, respectively.

Once the equation of state (EoS), $P(\varepsilon)$, is specified, the stable configurations of a NS then can be obtained from the well known hydrostatic equilibrium TOV equations as in our previous work~\cite{Li11y,Li15q,Li14y,Li12,Li10k,Li08q,Li06k}:
\beq\label{tov1:eps}
\frac{dP(r)}{dr}&=&-\frac{Gm(r)\varepsilon(r)}{r^{2}}\frac{\Big[1+\frac{P(r)}{\varepsilon(r)}\Big]\Big[1+\frac{4\pi r^{3}P(r)}{m(r)}\Big]}
 {1-\frac{2Gm(r)}{r}},\\
\frac{dm(r)}{dr}&=&4\pi r^{2}\varepsilon(r),
\eeq
where $P(r)$ is the pressure of the star at radius $r$, and $m(r)$ is the total star mass inside a sphere of radius $r$ ($G$ is the gravitational constant).

\section{results and discussion}

\begin{figure}
\vspace{0.3cm}
{\centering
\resizebox*{0.48\textwidth}{0.3\textheight}
{\includegraphics{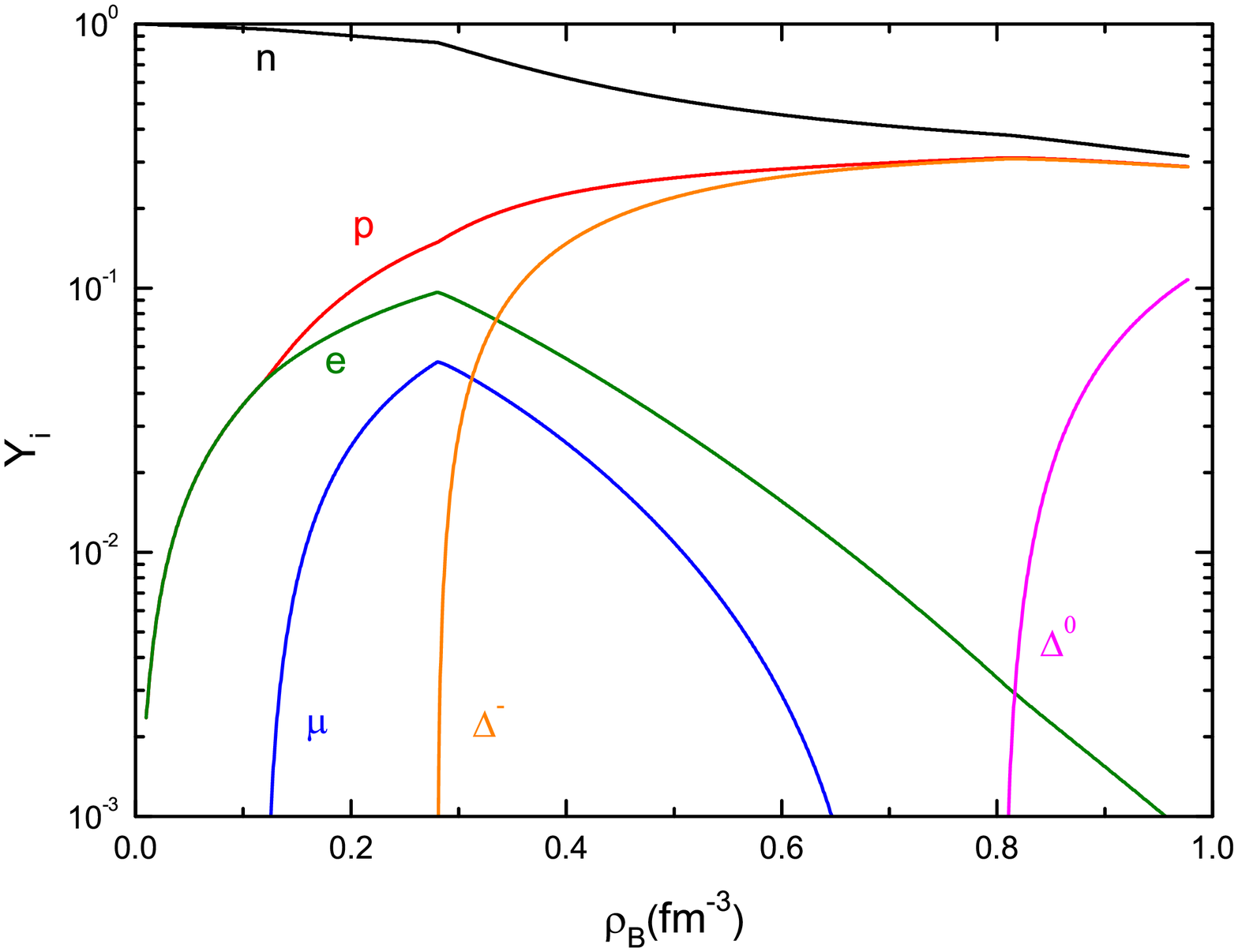}}
\par}
\caption{\small(Color online) Relative fraction $Y_i=\rho_i/\rho_B$ of $N\Delta e\mu$ matter as a function of the baryon density $\rho_B$ with the representative parameter set (PKO1) and fixed meson coupling of $(x_\sigma, x_\omega, x_{\rho}) = (0.8, 1.0, 0.5)$.}\label{fig3}
\end{figure}

We begin this section with the sensitivity study of the uncertain $\Delta$-meson couplings with one representative model parameter set (PKO1~\cite{long06plb}). As mentioned in the introduction, the calculations are done by varying the density-dependence features of $\Delta$ couplings with $\sigma, \omega, \rho$ mesons in reasonable ranges. The resulting $\Delta$ critical densities $\rho^{\rm crit}_{\Delta}/\rho_0$ are shown in Fig.~1 as functions of the presently-unknown $x_\rho$ in the range of $0.5-3.0$, for three cases of $(x_\sigma, x_\omega)$ of $ (0.8, 1.0), (1.0, 1.0), (1.0, 1.2)$. Previous calculations using RMF~\cite{li15prc} with $(x_\sigma, x_\omega) = (1.0, 1.0)$ are also shown for comparison.

We first notice a similarly linear $x_\rho$ dependence for DDRHF and RMF, namely, the larger $x_\rho$, the larger $\rho^{\rm crit}_{\Delta}/\rho_0$. However, due to remarkable adjustment of the coupling strengthes introduced by the density dependence in the present study, two main differences are present: One is an even earlier appearance of $\Delta$-isobars, at $\rho^{\rm crit}_{\Delta}/\rho_0 = 1.45$, to be compared with $\rho^{\rm crit}_{\Delta}/\rho_0  = 2.1$ with RMF~\cite{li15prc}, for the same choice of meson coupling of $(x_\sigma, x_\omega, x_{\rho}) = (1.0, 1.0, 1.0)$; The other is a weaker $x_\rho$ dependence, which is not subject to the modifications of $x_\sigma, x_\omega$ as one can notice from all three solid curves in Fig.~1. The increase of $x_\sigma ( x_\omega)$ leads to an earlier/later $\Delta$ appearance, which can be attributed to the attraction/repulsive nature of the $\sigma (\omega)$ coupling with $\Delta$-isobars in the dominate Hartree contributions.

\begin{figure}
\vspace{0.3cm}
{\centering
\resizebox*{0.48\textwidth}{0.3\textheight}
{\includegraphics{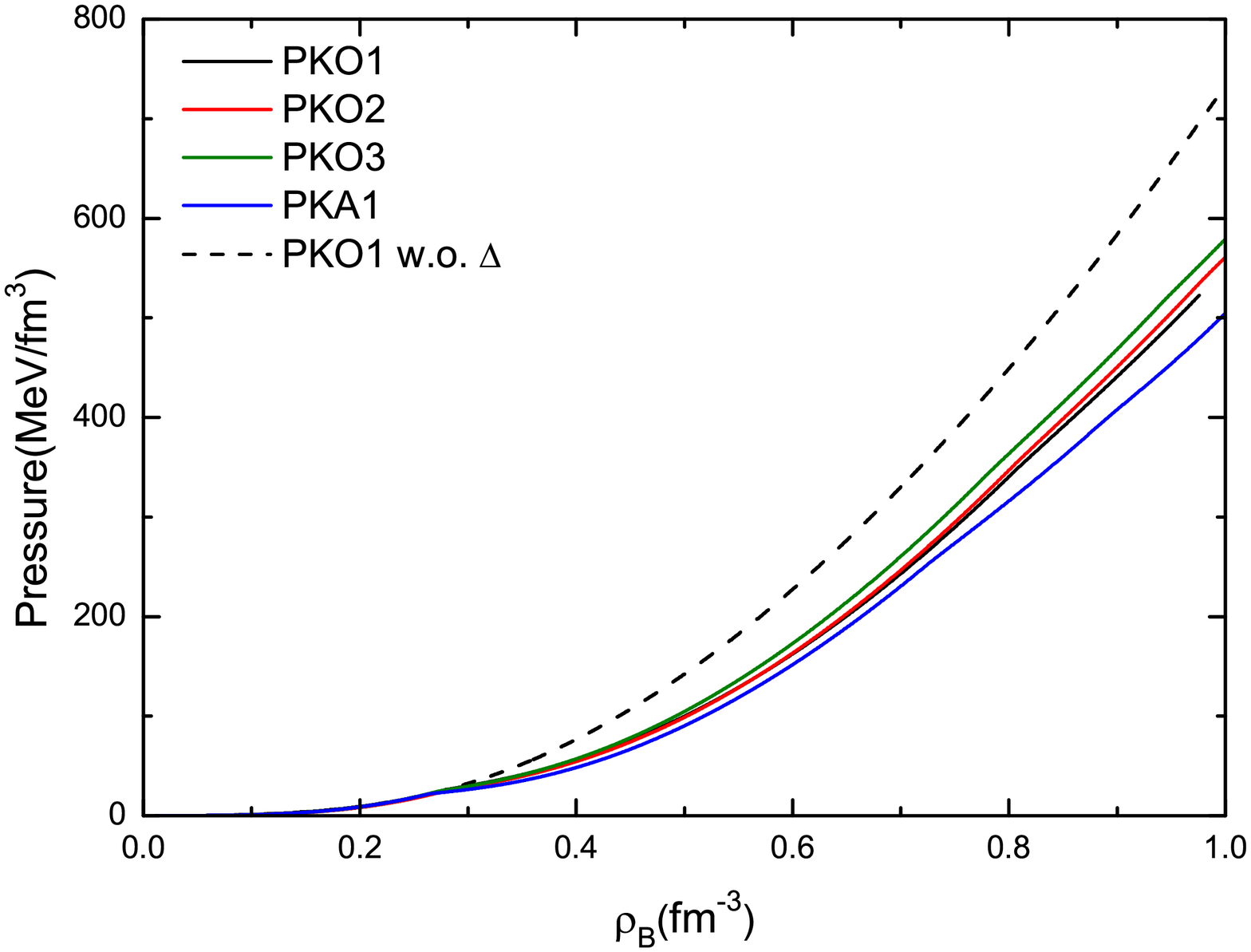}}
\par}
\caption{\small(Color online) Total pressure $P$ for $N\Delta e\mu$ matter as a function of the baryon density $\rho_B$ with all four available DDRHF parameter sets (PKO1, PKO2, PKO3, PKA1), at fixed meson-$\Delta$ coupling of $(x_\sigma, x_\omega, x_{\rho}) = (0.8, 1.0, 0.5)$. The results without $\Delta$-isobars with PKO1 are also shown for comparison.}\label{fig4}
\end{figure}

In Fig.~1, very similar $\Delta$ critical density is found for RMF at $(x_\sigma, x_\omega, x_{\rho}) = (1.0, 1.0, 0.5)$, and DDRHF at $(x_\sigma, x_\omega, x_{\rho}) = (0.8, 1.0, 0.5)$. Therefore, we do the following calculations mainly using one fixed meson coupling of $(x_\sigma, x_\omega, x_{\rho}) = (0.8, 1.0, 0.5)$ with the same representative PKO1 parameter set. We mention here that the choice of $x_\sigma = g_{\sigma\Delta} (\rho_B)/g_{\sigma N}(\rho_B)=0.8$ corresponds to a $20\%$ relaxation~\cite{rmf99npa} of the QCD result at $\rho_B=\rho_0$~\cite{qcd95}.

Fig.~2 shows the baryonic effective masses of Eq.~(34). For the study of $\Delta$-isobar effects, the results for $Ne\mu$ matter in the left panel are compared to those of $N\Delta e\mu$ matter in the right panel. All baryonic effective mass $M^*$ decrease with density because of the attractive $\sigma$ field for nucleons and $\Delta$-isobars. A slightly lower value for $M^*_n$ than $M^*_p$ in both panels is due to its larger fraction in the matter. Similarly, we can tell the merging sequence for $\Delta$-isobars as $\Delta^{-},\Delta^{0},\Delta^{+},\Delta^{++}$, following the requirements of chemical equilibrium of Eqs.~(34-35). With $\Delta^{-}$ first appearing at $\rho_B\sim0.28$ fm$^{-3}$, the proton population increases under the condition of charge neutrality, then $M^*_p$ decreases to be close to $M^*_n$ in the right panel for $N\Delta e\mu$ matter.

The detailed composition of $N\Delta e\mu$ matter is given in Fig.~3 as a function of the baryonic density. It is clear that the $\Delta$-isobars appear in the same sequence expected in the $M^*$ study in Fig.~2. After its early presence, $\Delta^-$ quickly replaces neutron and achieves a comparable (even higher) population with proton at high densities. It is a nature consequence of the interplay between the baryon number conservation and the charge neutrality.

\begin{figure}
\vspace{0.3cm}
{\centering
\resizebox*{0.48\textwidth}{0.3\textheight}
{\includegraphics{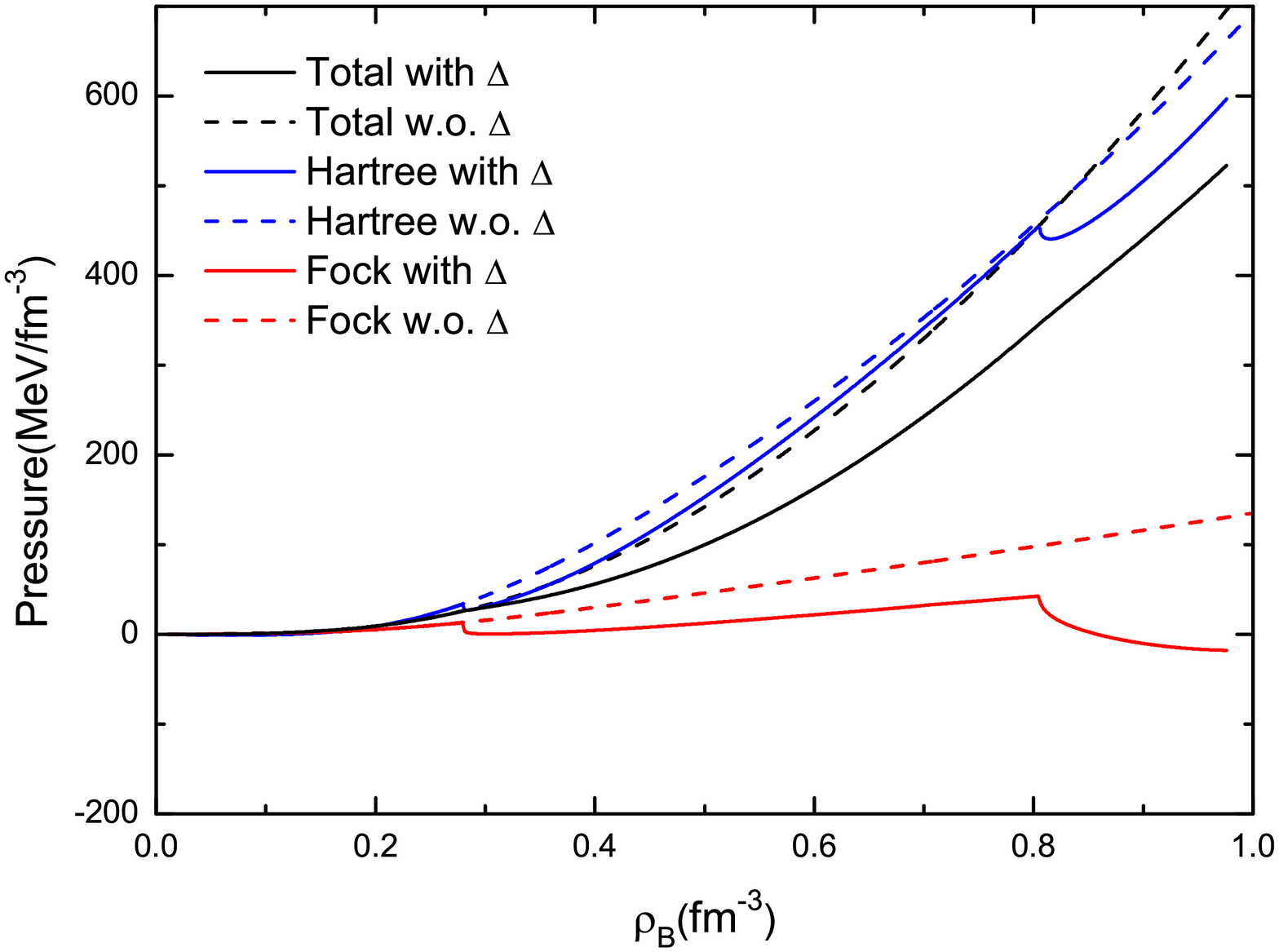}}
\par}
\caption{\small(Color online) Total pressure $P$ as a function of the baryon density $\rho_B$ for both $Ne\mu$ matter (in dashed lines) and $N\Delta e\mu$ matter (in solid lines). The corresponding contributions from the Hartree channels (in blue) and Fock channels (in red) also shown. The calculations are done for one representative parameter set (PKO1) and at fixed meson coupling of $(x_\sigma, x_\omega, x_{\rho}) = (0.8, 1.0, 0.5)$.}\label{fig5}
\end{figure}

The resulting EoSs are then presented in Fig.~4. Here all four available DDRHF parameter sets (PKO1, PKO2, PKO3~\cite{long06plb}, and PKA1~\cite{long08el}) are employed and compared. We see that at the $\Delta^-$ threshold density around $0.28$ fm$^{-3}$, the pressure curves of $N\Delta e\mu$ matter start to differ from (actually be softer than) those of $Ne\mu$ matter. The modification of the $\Delta^-$ threshold density due to different model parameter employed is shown to be quite limited (only $\leq5.99\%$). For example it is 0.2805 fm$^{-3}$, 0.2775 fm$^{-3}$, 0.2715 fm$^{-3}$, 0.2640 fm$^{-3}$ for PKO1, PKO2, PKO3, PKA1, respectively. This weak-dependence feature was not observed in previous RMF calculations~\cite{guo03prc}. In particular, as shown in Fig.~2 of Ref.~\cite{guo03prc}, to substitute the parameter set from NL1 to TM1, the modification in the $\Delta$ threshold in the nonlinear Walecka model is $\sim37.6\%$; Similarly, a change from T1 to T3 for a chiral hadronic model resulted in a corresponding modification of $\sim28.9\%$. In addition, we notice that in the calculations of the nonlinear Walecka model~\cite{guo03prc}, the $\Delta$-excited matter is found to be the ground state of nuclear matter instead of normal $Ne\mu$ matter, which might not be physically valid. We also mention here that a much delayed appearance of $\Delta$-isobars (the earliest of $3.22\rho_0$ for T1 parameter set) was found in the calculation of the chiral hadronic model~\cite{guo03prc}, even with an enhanced $\sigma$-meson coupling of $x_{\sigma}=1.39$. The last point demonstrated in Fig.~4 for DDRHF is that, with increasing density, the PKA1 (PKO3) parameter set provides as the softest (stiffest) EoS, with the PKO1 and PKO2 cases in the middle. Moreover, the weak-dependence feature continues at high densities, with the pressure only lifting $\sim12.8\%$ from softest PKA1 to the stiffest PKO3 at the highest density plotted here (1.0 fm$^{-3}$).

\begin{figure}
\vspace{0.3cm}
{\centering
\resizebox*{0.48\textwidth}{0.3\textheight}
{\includegraphics{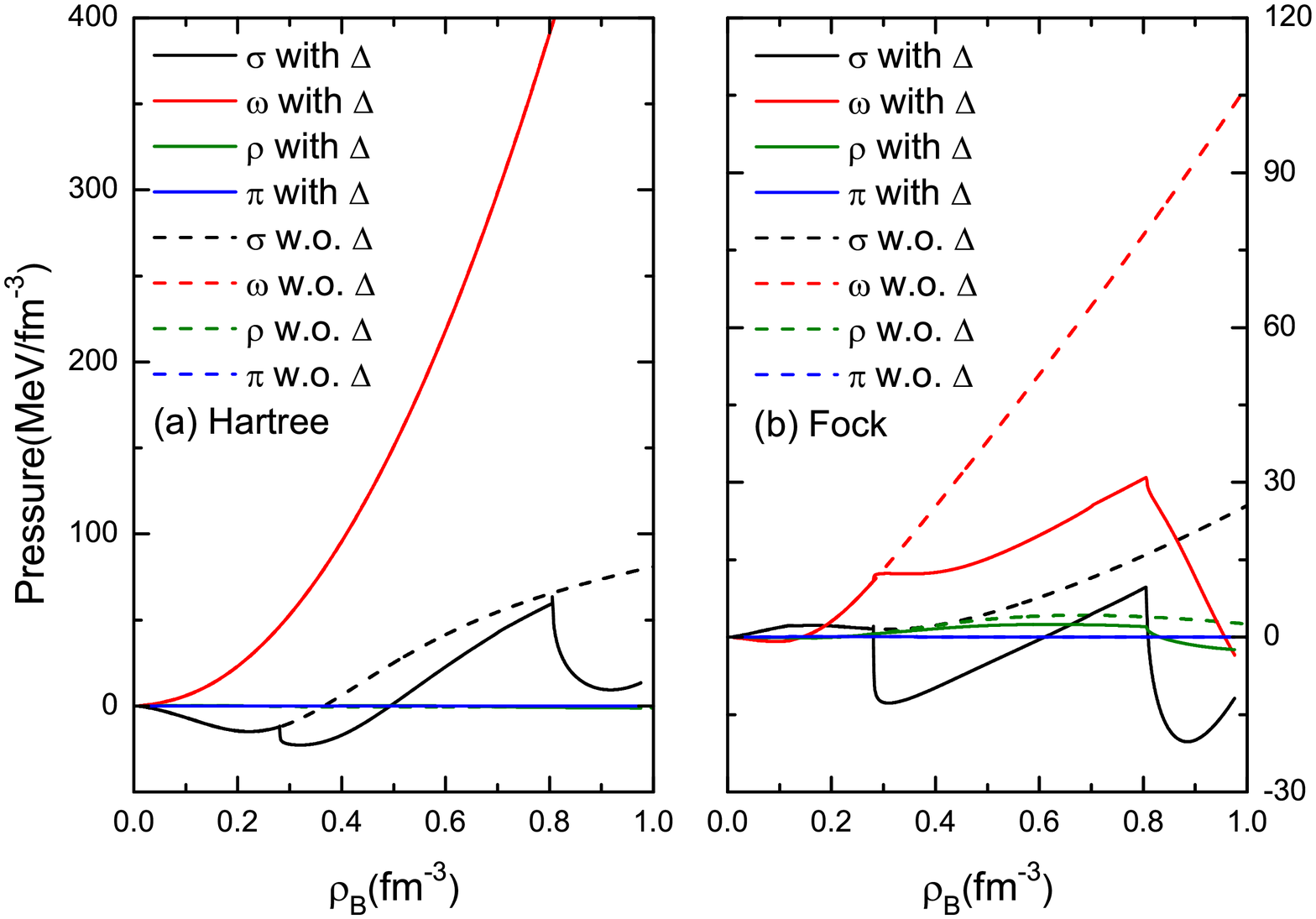}}
\par}
\caption{\small(Color online) Pressure contributions from different mesons for both the Hartree channels (left panel) and Fock channels (right panel), with (in solid lines) or without (in dashed lines) $\Delta$-isobars. The calculations are done for one representative parameter set (PKO1) and at fixed meson coupling of $(x_\sigma, x_\omega, x_{\rho}) = (0.8, 1.0, 0.5)$.}\label{fig6}
\end{figure}

In the following two figures, we continue to use PKO1 as the representative parameter set to discuss in details the separate contributions from various meson terms for the EoSs. The meson couplings are also fixed to be $(x_\sigma, x_\omega, x_{\rho}) = (0.8, 1.0, 0.5)$.

In Fig.~5, the total pressures of both $Ne\mu$ matter and $N\Delta e\mu$ matter are plotted together with the corresponding contributions from the Hartree terms and Fock terms. One can immediately notice that there are discontinuous in separate Hartree/Fock contributions at threshold densities of $\Delta$-isobars, as a result of the third term in the right hand side of Eq.~(40). Both the Hartree channel and Fock channel become less repulsive once a new type of $\Delta$-isobars appears. Also, for both two cases of with and without $\Delta$-isobars, the total pressures are dominated by the Hartree channels, which stay repulsive at high densities. However, the Fock contribution could be decreased to be attractive at high densities (around $0.86$ fm$^{-3}$) after the presence of $\Delta^-$ and $\Delta^0$. One may then conclude that the $\Delta$ softening on the EoS is largely due to a reduced Fock contribution introduced by $\Delta$-isobars.

To further understand this point, we separate the contributions from different mesons in Fig.~6, where the Hartree/Fock part is presented in the left/right panel. We see that the contributions from the isovector mesons ($\pi$, $\rho$) are much smaller than those of the isoscalar mesons ($\sigma$, $\omega$). Also, although at $\rho_B=\rho_0$ the $\omega$ contribution is comparable with the $\sigma$ contribution, its increasing with density is more pronounced than that of the $\sigma$ contribution. As a result, the pressures in both the Hartree part and the Fock part are dominated by the repulsive $\omega$ couplings to both nucleons and $\Delta$-isobars. In the right panel for the Fock part, we can easily observe strong decreasing of the repulsive $\sigma, \omega$ couplings to the $\Delta$-isobars. This is the main reason why the Fock term contributes a negative pressure at high densities in Fig.~5. In particular, the reducing $\Delta$-$\omega$ Fock coupling is enhanced once a new type of $\Delta$-isobars appears. A similar conclusion has been previously found when Long et al.~\cite{long12prc} studied the $\Lambda$-matter for the $\Lambda$-$\omega$ coupling within DDRHF.

\begin{figure}
\vspace{0.3cm}
{\centering
\resizebox*{0.48\textwidth}{0.3\textheight}
{\includegraphics{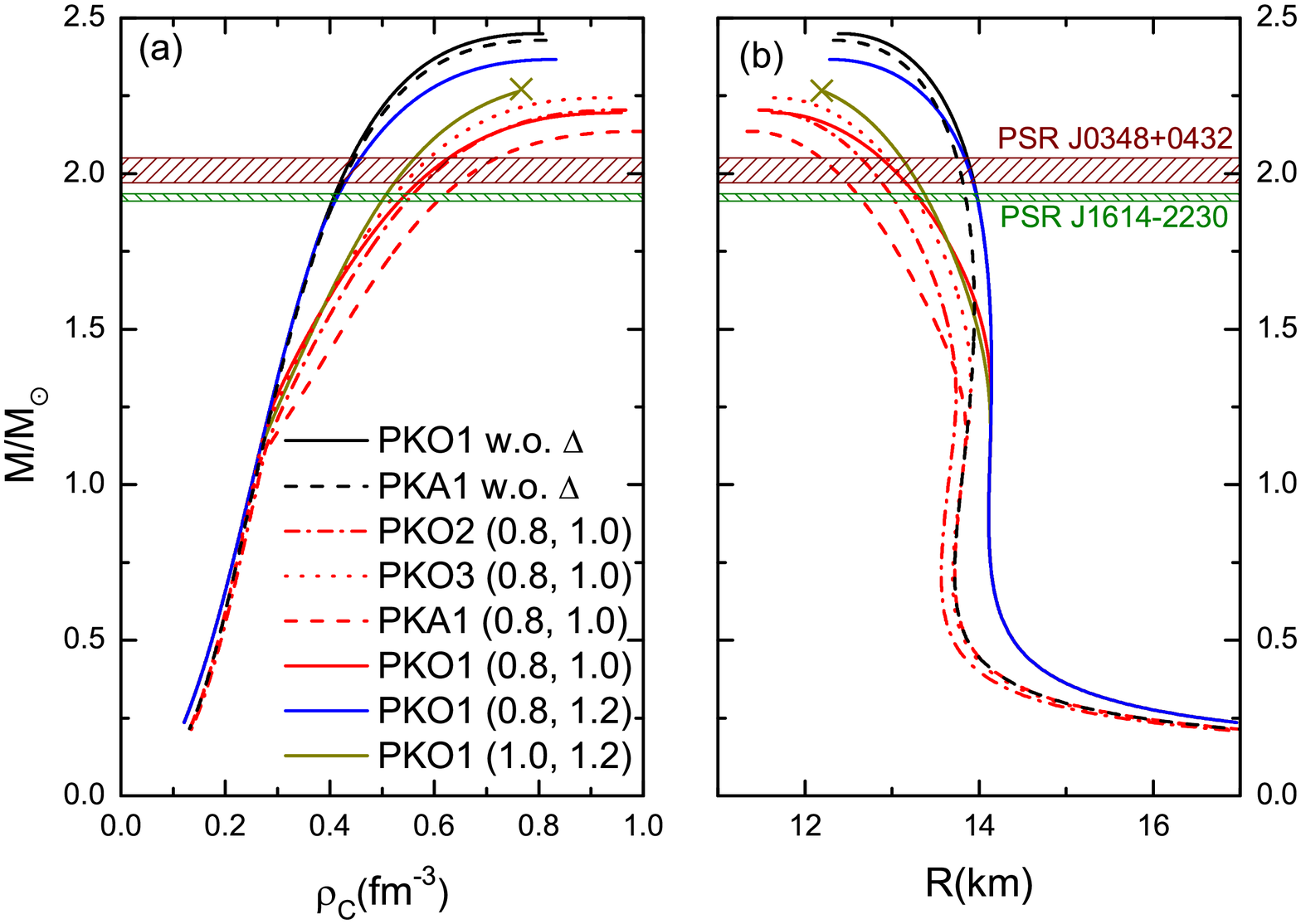}}
\par}
\caption{\small(Color online) Gravitational mass $M$ as a function of the central density $\rho_c$ (the stellar radius $R$) in the left (right) panel. The calculations with PKO1 parameter set are done for three cases of $\Delta$-meson coupling of $(x_\sigma, x_\omega) = (0.8, 1.0), (1.0, 1.0), (1.0, 1.2)$ at fixed $x_\rho=0.5$. The calculations with PKA1, PKO2, PKO3 parameter sets are done for one case of $\Delta$-meson coupling of $(x_\sigma, x_\omega) = (0.8, 1.0)$. The results without $\Delta$-isobars for PKO1 and PKA1 are also shown for comparisons, together with two recent massive stars: PSR J1614-2230~\cite{2solar10,2solar11} and PSR J0348+0432~\cite{2solar2}. The  cross in the $(x_\sigma, x_\omega) = (1.0, 1.2)$ curve is the point beyond which no physical solution is found.}\label{fig7}
\end{figure}

Finally the NS structures are calculated based on the EoSs obtained in the present study. The results are shown in Fig.~7, where the star's gravitational mass is shown as a function of the central density (the stellar radius) in the left (right) panel. All four available DDRHF parameter sets are employed and the calculations are done for three cases of $\Delta$-meson coupling of $(x_\sigma, x_\omega) = (0.8, 1.0), (1.0, 1.0), (1.0, 1.2)$. We see that the introducing of $\Delta$-isobars leads to a decrease of the predicted maximum mass for NSs, as expected from adding a new degree of freedom to the system. For example at $(x_\sigma, x_\omega) = (0.8, 1.0)$, a maximum mass of 2.50$M_{\odot}$ (2.43$M_{\odot}$) without $\Delta$-isobars is lowered to 2.24$M_{\odot}$ (2.14$M_{\odot}$) for PKO3 (PKA1). The mass (radius) difference between PKO3 and PKA1 are $1.45\%$ ($0.10\%$) without $\Delta$-isobars, $2.46\%$ ($1.34\%$) with $\Delta$-isobars, although the symmetry energy parameters differ considerably, i.e., $(E_{\rm sym}, L)= (32.99$ MeV$, 83.00$ MeV) for PKO3, and $(E_{\rm sym}, L)= (36.02$ MeV$, 103.50$ MeV) for PKA1. These weak sensitivities of star predictions on symmetry energy parameters (especially $L$) were demonstrated before~\cite{long12prc} in the case without $\Delta$-isobars. Model studies with RMF using constant meson couplings, however, show much larger sensitivities~\cite{long12prc}. We then conclude here that the introduction of density-dependence for meson couplings could lead to concentrated results on predictions of star properties.

Moreover, the maximum mass increases/decreases with increasing $x_\omega/x_\sigma$, as expected from the stiffened/softened EoS obtained above. To take PKO1 as an example, a maximum mass of 2.45$M_{\odot}$ without $\Delta$-isobars is lowered to 2.37$M_{\odot}$, 2.19$M_{\odot}$ corresponding to $(x_\sigma, x_\omega) = (0.8, 1.2), (0.8, 1.0)$. The radius also smaller, from $12.38$ km down to $12.28$ km, $11.62$ km, respectively. It is important to point out here that lower $L$ values are suggested~\cite{apj13a,apj14,apj13b} by both nuclear experiments, chiral effective theory, and some neutron star radii. New DDRHF parameter sets should be proposed in line with them.
The cross in the $(x_\sigma, x_\omega) = (1.0, 1.2)$ curve is the point where the $\Delta$ population is close to that of nucleons, and no physical solution is found after that point. In that case, the HF potential for spin-$3/2$ $\Delta$-isobars can not be self-consistently obtained, even the in-medium $\Delta$ effective mass is still far from zero.

%-------------------------------------------------------------------------------
\section{Conclusions}

Summarizing, we have extended the DDRHF theory to include the $\Delta$-isobars for the interest of dense matter and NSs. For this purpose, we solve the Rarita-Schwinger equation for spin-3/2 particle with the $\Delta$ self-energy determined self-consistently. Four mesons (the isoscalar $\sigma$ and $\omega$ as well as isovector $\rho$ and $\pi$) are included for producing interaction between the baryons. All four available parameter sets (PKO1, PKO2, PKO3, PKA1) are employed to explore the model utilization. Also, the calculations are done with various choices of the uncertain meson-$\Delta$ couplings following the constrains
reported in the various analysis of experimental data.

We found that $\Delta$-isobars appear early in the nuclear matter, soften the EoS of NS matter, and reduce the NSs' maximum mass. We observed a controlled behaviour of the results, with respect to the change of the model parameter set and different meson-$\Delta$ couplings. In particular, the $\Delta$ softening of the EoS is due to the large decrease in Fock channel, mainly from the isoscalar
mesons. On the other hand, the pion contributions are found to be negligibly small. Finally, we conclude that within DDRHF, a NS with $\Delta$-isobars in its core can be as heavy as the two recent massive stars whose masses are accurately measured, with a slightly smaller radius than the corresponding normal NSs.

\section{Acknowledgments}
 This work was supported by the National Natural Science Foundation of China (Grants Nos. 11405090 and U1431107).

\section{Appendix~A}
\widetext
The vertices $\Gamma_\lambda^{ab}(1,2)$ between mesons and baryons are:
\beq
& \Gamma_\sigma^{ab}(1,2)&= -g_{\sigma a}g_{\sigma b}, \\
& \Gamma_\omega^{ab}(1,2)& = g_{\omega a}g_{\omega b}\gamma^\mu(1)\gamma_\mu(2), \\
& \Gamma_\pi^{ab}(1,2)& = -\frac{f_{\pi a}f_{\pi b}}{m_\pi^2}(\gamma_5\slashed{q}\bm{\tau}_b)_1(\gamma_5\slashed{q}\bm{\tau}_a)_2, \\
& \Gamma_{\rho_V}^{ab}(1,2)& = g_{\rho a}g_{\rho b}\gamma^\mu(1)\bm{\tau}_b(1) \gamma_\mu(2)\bm{\tau}_a(2), \\
& \Gamma_{\rho_T}^{ab}(1,2)& = \frac{f_{\rho a}f_{\rho b}}{4M_a M_b} (q_\lambda \sigma^{\mu\lambda}\bm{\tau}_b)_1 (q^\nu \sigma_{\mu\nu}\bm{\tau}_a)_2, \\
& \Gamma_{\rho_{VT}}^{ab}(1,2)& = \frac{if_{\rho a}g_{\rho b}}{2M_ a} \gamma^\mu(1)\bm{\tau}_b(1) q^\nu\sigma_{\mu\nu}(2)\bm{\tau}_a(2)
- \frac{if_{\rho b}g_{\rho a}}{2M_b} q_\nu\sigma^{\mu\nu}(1)\bm{\tau}_b(1)\gamma_\mu(2)\bm{\tau}_a(2),
\eeq
where $a,b$ denote nucleon or $\Delta$-isobars.

\section{Appendix~B}
Here we calculate the energy density of the system. we first write the Hamiltonian density $H$ (Eq.~(3)) in terms of $u(p,\kappa)$ and $u^\alpha(p,\kappa)$, with the corresponding kinetic energy term $T$ and the potential term $V$ as
\beq
% ---------------------------------------------kinetic term-----------------------------------------------------------
T &=& \sum_{\bm{p},\kappa_1,\kappa_2}\bar{u}_N(\bm{p},\kappa_1)(\bm{\gamma} \cdot\bm{p} + M_N) u_N(\bm{p},\kappa_2)b_{\bm{p},\kappa_1}^{N\dagger} b_{\bm{p},\kappa_2}^N + \sum_{\bm{p},\kappa_1,\kappa_2}\bar{u}_{\Delta\alpha}(\bm{p},\kappa_1)(\bm{\gamma} \cdot\bm{p} + M_\Delta) u_\Delta^\alpha(\bm{p},\kappa_2)b_{\bm{p},\kappa_1}^{\Delta\dagger} b_{\bm{p},\kappa_2}^\Delta,\nn
% -----------------------------------------------NN part--------------------------------------------------------------
V &= & \frac{1}{2}\sum_\lambda\sum_{\bm{q},\bm{p}_1,\bm{p}_2}\sum_{\substack{\kappa_1,\kappa_2, \\ \kappa_3,\kappa_4}}
\bar{u}_N(\bm{p}_1,\kappa_1)\bar{u}_N(\bm{p}_2,\kappa_2)\frac{\Gamma_\lambda^{NN}}{m_\lambda^2+\bm{q}^2} u_N(\bm{p}_2+\bm{q},\kappa_3)u_N(\bm{p}_1-\bm{q},\kappa_4) b_{\bm{p}_1,\kappa_1}^{N\dagger}b_{\bm{p}_2,\kappa_2}^{N\dagger}b_{\bm{p}_2+\bm{q},\kappa_3}^N b_{\bm{p}_1-\bm{q},\kappa_4}^N \nonumber \\
% ---------------------------------------------N Delta part-----------------------------------------------------------
&& + \frac{1}{2}\sum_\lambda\sum_{\bm{q},\bm{p}_1,\bm{p}_2}\sum_{\substack{\kappa_1,\kappa_2, \\ \kappa_3,\kappa_4}}
\bar{u}_N(\bm{p}_1,\kappa_1)\bar{u}_{\Delta\beta}(\bm{p}_2,\kappa_2)\frac{\Gamma_\lambda^{N\Delta}}{m_\lambda^2+\bm{q}^2} u_\Delta^\beta(\bm{p}_2+\bm{q},\kappa_3)u_N(\bm{p}_1-\bm{q},\kappa_4) b_{\bm{p}_1,\kappa_1}^{N\dagger}b_{\bm{p}_2,\kappa_2}^{\Delta\dagger}b_{\bm{p}_2+\bm{q},\kappa_3}^\Delta b_{\bm{p}_1-\bm{q},\kappa_4}^N \nonumber \\
% ---------------------------------------------Delta N part-----------------------------------------------------------
&& + \frac{1}{2}\sum_\lambda\sum_{\bm{q},\bm{p}_1,\bm{p}_2}\sum_{\substack{\kappa_1,\kappa_2, \\ \kappa_3,\kappa_4}}
\bar{u}_{\Delta\alpha}(\bm{p}_1,\kappa_1)\bar{u}_N(\bm{p}_2,\kappa_2)\frac{\Gamma_\lambda^{\Delta N}}{m_\lambda^2+\bm{q}^2} u_N(\bm{p}_2+\bm{q},\kappa_3)u_\Delta^\alpha(\bm{p}_1-\bm{q},\kappa_4) b_{\bm{p}_1,\kappa_1}^{\Delta\dagger}b_{\bm{p}_2,\kappa_2}^{N\dagger}b_{\bm{p}_2+\bm{q},\kappa_3}^N b_{\bm{p}_1-\bm{q},\kappa_4}^\Delta \nonumber \\
% -------------------------------------------Delta Delta part---------------------------------------------------------
&& + \frac{1}{2}\sum_\lambda\sum_{\bm{q},\bm{p}_1,\bm{p}_2}\sum_{\substack{\kappa_1,\kappa_2, \\ \kappa_3,\kappa_4}}
\bar{u}_{\Delta\alpha}(\bm{p}_1,\kappa_1)\bar{u}_{\Delta\beta}(\bm{p}_2,\kappa_2)\frac{\Gamma_\lambda^{\Delta\Delta}} {m_\lambda^2+\bm{q}^2} u_\Delta^\beta(\bm{p}_2+\bm{q},\kappa_3)u_\Delta^\alpha(\bm{p}_1-\bm{q},\kappa_4) b_{\bm{p}_1,\kappa_1}^{\Delta\dagger}b_{\bm{p}_2,\kappa_2}^{\Delta\dagger}b_{\bm{p}_2+\bm{q},\kappa_3}^\Delta b_{\bm{p}_1-\bm{q},\kappa_4}^\Delta.\nonumber \\
\eeq

With Hartree-Fock wave function in Eq.~(31), the kinetic term is calculated as:
\beq
% ===================================================N term=============================================================
\langle T\rangle =\sum_{k=n,p} \frac{1}{\pi^2}\int_0^{p_{F,k}}p^2dp(M_N\hat{M}_k + p\hat{P}_k) + \sum_{\substack{i=\Delta^-,\Delta^0, \\ \Delta^+, \Delta^{++}}}\frac{2}{\pi^2}\int_0^{p_{F,i}}p^2dp(M_\Delta\hat{M}_i + p\hat{P}_i).
\eeq
The potential term can be decomposed into the direct term $\langle V_D\rangle$ and the exchange term $\langle V_D\rangle$, respectively. For the direct potential term:
\beq
% ==================================================NN term=============================================================
\langle V_D\rangle &=& -\frac{1}{2}\frac{g_{\sigma_N}^2}{m_\sigma^2}\rho_{S_N}^2
+ \frac{1}{2}\frac{g_{\omega_N}^2}{m_\omega^2}\rho_{B_N}^2 + \frac{1}{2}\frac{g_{\rho_N}^2}{m_\rho^2}\rho_{B_{N3}}^2
% =================================================ND+DN term===========================================================
% they are same
+ \frac{g_{\sigma_N}g_{\sigma_\Delta}}{m_\sigma^2}\rho_{S_N}\rho_{S_\Delta}
- \frac{g_{\omega_N}g_{\omega_\Delta}}{m_\omega^2}\rho_{B_N}\rho_{B_\Delta}
- \frac{g_{\rho_N}g_{\rho_\Delta}}{m_\rho^2}\rho_{B_{N3}}\rho_{B_{\Delta3}}\nn
% ==================================================DD term=============================================================
&&- \frac{1}{2}\frac{g_{\sigma_\Delta}^2}{m_\sigma^2}\rho_{S_\Delta}^2
+ \frac{1}{2}\frac{g_{\omega_\Delta}^2}{m_\omega^2}\rho_{B_\Delta}^2
+ \frac{1}{2}\frac{g_{\rho_\Delta}^2}{m_\rho^2}\rho_{B_{\Delta3}}^2,
\eeq
where, the densities of nucleon and $\Delta$-isobars are defined as
\beq
 \rho_{B_N} = \rho_{B_n} + \rho_{B_p}, &~&\rho_{B_{\Delta}} = \rho_{B^{++}} + \rho_{B^+} + \rho_{B^0} + \rho_{B^-},\\
\rho_{B_{N_k}} = \frac{1}{\pi^2}\int_0^{p_{F,k}}p^2dp, &~&
 \rho_{B_{\Delta_i}} = \frac{2}{\pi^2}\int_0^{p_{F,i}}p^2dp,\\
 \rho_{S_N} = \rho_{S_n} + \rho_{S_p}, &~& \rho_{S_\Delta} = \rho_{S^{++}} + \rho_{S^+} + \rho_{S^0} + \rho_{S^-},\\
\rho_{S_{N_k}} = \frac{1}{\pi^2}\int_0^{p_{F,k}}p^2dp \hat{M}(p), &~&
 \rho_{S_{\Delta_i}} = \frac{2}{\pi^2}\int_0^{p_{F,i}}p^2dp \hat{M}(p),\\
\rho_{B_{N3}} = \rho_{B_p} - \rho_{B_n}, &~&
 \rho_{B_{\Delta3}} = 3\rho_{B^{++}} + \rho_{B^+} - \rho_{B^0} - 3\rho_{B^-}
\eeq
%------------------------------------------------
\begin{table}\label{tab1}
\caption{Functions $A_\lambda$, $B_\lambda$, $C_\lambda$ of $\lambda = \sigma, \omega, \pi, \rho_T, \rho_V$ couplings for calculating $\langle V_{E_N}\rangle$.}
\begin{center}
\begin{tabular}{cccc} \hline
$\lambda$ & $A_\lambda$ & $B_\lambda$ & $C_\lambda$ \\ \hline
$\sigma$ & $-g_\sigma^2 \theta_\sigma$ & $-g_\sigma^2 \theta_\sigma$ & $g_\sigma^2 \phi_\sigma$ \\
$\omega$ & $-2g_\omega^2 \theta_\omega$ & $4g_\omega^2 \theta_\omega$ & $2g_\omega^2 \phi_\omega$ \\
$\rho_V$ & $-2g_\rho^2 \theta_\rho$ & $4g_\rho^2 \theta_\rho$ & $2g_\rho^2 \phi_\rho$ \\
$\pi$ & $\left(\cfrac{f_\pi}{m_\pi}\right)^2 m_\pi^2 \theta_\pi$ &
$\left(\cfrac{f_\pi}{m_\pi}\right)^2 m_\pi^2 \theta_\pi$ &
$\left(\cfrac{f_\pi}{m_\pi}\right)^2 [-(p^2+\mathop{p'}^2)\phi_\pi + 2\mathop{pp'}\theta_\pi]$ \\
$\rho_T$ & $\left(\cfrac{f_\rho}{2M}\right)^2 m_\rho^2 \theta_\rho$ &
$3\left(\cfrac{f_\rho}{2M}\right)^2 m_\rho^2 \theta_\rho$ &
$\left(\cfrac{f_\rho}{2M}\right)^2 [-2(p^2+\mathop{p'}^2-m_\rho^2 /2)\phi_\pi + 4\mathop{pp'}\theta_\rho]$ \\\hline
$\rho_{VT}$ & & $D=12\cfrac{f_\rho g_\rho}{2M}(-\mathop{p'}\theta_\rho + p\phi_\rho)$ \\ \hline
\end{tabular}
\end{center}
\end{table}
The calculation of exchange contribution from nucleons and $\Delta$-isobars can be changed into trace calculations as
\beq
\langle V_{E_N}\rangle = \frac{1}{2}\sum_\lambda\sum_{\tau_1,\tau_2}\sum_{p_1,p_2}\frac{1}{m_\lambda^2 + \bm{q}^2} \text{Tr}[D_N(p_1,\tau_1)\Gamma^{NN}_\lambda(2)
D_N(p_2,\tau_2)\Gamma^{NN}_\lambda(1)],\\
\langle V_{E_\Delta}\rangle = \frac{1}{2}\sum_\lambda\sum_{\tau_1,\tau_2}\sum_{p_1,p_2}\frac{1}{m_\lambda^2+\bm{q}^2}\text{Tr}
[S_{\Delta}^{\alpha\beta}(p_1,\tau_1)\Gamma^{\Delta\Delta}_\lambda(2) S_{\Delta,{\alpha\beta}}(p_2,\tau_2)\Gamma^{\Delta\Delta}_\lambda(1)].
\eeq
Here $D_N(p,\tau)$ is the propagator of nucleons:
\beq
D_N(p,\tau) = \sum_s u(p,s,\tau)\bar{u}(p,s,\tau) = \frac{\slashed{p}\ast + M^\ast}{2E^\ast},
\eeq
and $S_\Delta^{\alpha\beta}(p,\tau)$ is the propagator of $\Delta$-isobars:
\beq
S_\Delta^{\alpha\beta}(p,\tau) = \sum_s u_\Delta^\alpha(p,s,\tau)\bar{u}_\Delta^\beta(p,s,\tau) = \frac{\slashed{p}^\ast + M^\ast}{2E^\ast}\left[-g^{\alpha\beta} + \frac{1}{3}\gamma^\alpha\gamma^\beta
+ \frac{2}{3M^{\ast2}}p^\alpha p^\beta - \frac{1}{3M^\ast}(p^{\alpha\ast}\gamma^\beta - p^{\beta\ast}\gamma^\alpha)\right]
\eeq
The nucleonic exchange terms of Eq.~(58) for symmetric nuclear matter were calculated in Ref.~\cite{bouyssy87prc}. We write the formula here for asymmetric  nuclear matter ($k,l$ denote neutron or proton):
\beq
\langle V_{E_N}\rangle = \frac{1}{(2\pi)^4}\sum_{k,l}\int_0^{p_{F,k}}pdp\int_0^{p_{F,l}}\mathop{p'dp'}\frac{1}{2}\left[\sum_\lambda \Omega_{kl}^\lambda \biggl(A_\lambda + B_\lambda\mathop{\hat{M}\hat{M}'} + C_\lambda\mathop{\hat{P}\hat{P}'}\biggr) + \frac{1}{2}\Omega_{kl} ^\rho \biggl(\mathop{\hat{P}\hat{M}'}D(\mathop{p,p'}) + \mathop{\hat{P}'\hat{M}}D(\mathop{p',p})\biggr)\right],
\eeq
where $\hat{P}(\mathop{p})\equiv \hat{P},~\hat{P}(\mathop{p'})\equiv \hat{P'}$ are used in order to shorten the notation. The matrix element $\Omega_{kl}^\lambda$ is related to the isospin part and will be evaluated in Appendix C. Functions $A_\lambda, B_\lambda, C_\lambda, D$ for $\lambda = \sigma, \omega, \pi, \rho_V, \rho_T, \rho_{TV}$ are shown in Table I. The functions in Table I are defined as
\beq
\theta_\lambda(\mathop{p,p'})&=&\ln\biggl[\frac{m_\lambda^2+(\mathop{p-p'})^2}{m_\lambda^2+(\mathop{p+p'})^2}\biggr],\\
\phi_\lambda(\mathop{p,p'})&=&\frac{\mathop{p^2 + p'^2} + m_\lambda^2}{\mathop{2pp'}}\theta_\lambda(\mathop{p,p'}) + 2.
\eeq
Now we turn to the $\Delta$ parts in Eq.~(59). We first do trace calculation for each meson, defining
\beq
R_\lambda \equiv \text{Tr}[S_{\Delta}^{\alpha\beta}(p_1,\tau_1)\Gamma^{\Delta\Delta}_\lambda(2) S_{\Delta,{\alpha\beta}}(p_2,\tau_2)\Gamma^{\Delta\Delta}_\lambda(1)].
\eeq
We have
\beq
&&R_\sigma = \frac{1}{9} g_{\sigma_\Delta}^2(-10b - 14a - 8\hat{a}a - 4\hat{a}^2a),\\
&&R_\omega  = \frac{1}{9}g_{\omega_\Delta}^2 (32b - 4a + 16\hat{a}a - 8\hat{a^2}a),\\
&&R_\pi= \frac{1}{9}\biggl(\frac{f_{\pi_\Delta}}{m_\pi}\biggr)^2 (-20c_1c_2 + 6ad + 6bd + 4c_1^2\frac{\hat{M}_2}{\hat{M}_1} + 4c_2^2\frac{\hat{M}_1}{\hat{M}_2} - 8\hat{a}^2c_1c_2 + 4\hat{a}^2ad + 4\hat{a}ad),\\
&&R_{\rho_V} = \frac{1}{9}g_{\rho_\Delta}^2 (32b - 4a + 16\hat{a}a - 8\hat{a^2}a),\\
&&R_{\rho_T}= \frac{1}{9}\biggl(\frac{f_{\rho_\Delta}}{2M_\Delta}\biggr)^2 (-24c_1c_2 + 6ad + 22bd - 16\hat{a}^2c_1c_2 + 4\hat{a}^2ad + 14\hat{a}ad),\\
&&R_{\rho_{TV}}= \frac{1}{9}\frac{f_{\rho_\Delta}g_{\rho_\Delta}}{2M_\Delta}(44\hat{c}_1 - 44\hat{c}_2 - 8\hat{a}\hat{c}_2 + 8\hat{a}\hat{c}_1 - 24\hat{a}^2\hat{c}_2 + 24\hat{a}^2\hat{c}_1)\hat{M}_1\hat{M}_2,
\eeq
 where $\hat{M}(\mathop{p_1})\equiv \hat{M_1},~\hat{M}(\mathop{p_2})\equiv \hat{M_2} $ have been used in order to shorten the notation.

The coefficients $\hat a,~a,~b,~c_1,~c_2,~\hat{c}_1,~\hat{c}_2$, $d$ are defined as:
\beq
\hat{a}&=&\frac{p^\ast_1\cdot p^\ast_2}{M_1^\ast M_2^\ast} = \frac{1}{\hat{M}(p_1)\hat{M}(p_2)} - \frac{\hat{P}(p_1) \hat{P}(p_2)}{\hat{M}(p_1)\hat{M}(p_2)} \cos\theta,\\
a &=& 1 - \hat{P}(p_1)\hat{P}(p_2)\cos \theta,\\
b &=& \hat{M}(p_1)\hat{M}(p_2),\\
c_1&=& \frac{p_1^\ast\cdot q}{E_1^\ast} = \hat{P}(p_1)p_2\cos \theta - \hat{P}(p_1)p_1, \\
c_2&=& \frac{p_2^\ast\cdot q}{E_1^\ast} = \hat{P}(p_2)p_2 - \hat{P}(p_2)p_1\cos \theta,\\
\hat{c}_1 &=& \frac{\hat{P}(p_1)}{\hat{M}(p_1)}p_2\cos \theta - \frac{\hat{P}(p_1)}{\hat{M}(p_1)}p_1,\\
\hat{c}_2 &=& \frac{\hat{P}(p_2)}{\hat{M}(p_2)}p_2 - \frac{\hat{P}(p_2)}{\hat{M}(p_2)}p_1\cos \theta,\\
d &=& \bm{q}^2.
\eeq
%========================================================================================================================
\begin{table*}
\tabcolsep 1pt
\caption{Functions $\hat{H}^m$ and $A_{\lambda}^m$ ($m = 1, 2 ...12$) of $\lambda = \sigma, \omega, \pi, \rho_T, \rho_V$ couplings for calculating $\langle V_{E_\Delta}\rangle$.}
\vspace*{-12pt}
\begin{center}
\def\temptablewidth{0.96\textwidth}
{\rule{\temptablewidth}{0.5pt}}
\begin{tabular*}{\temptablewidth}{@{\extracolsep{\fill}}|c|c|c|c|c|c|c|}
\multicolumn{2}{|c|}{\diagbox{$\hat{H}^m$}{$A_{\lambda}^m$}} & $A_\sigma^m$ & $A_\omega^m$ & $A_\pi^m$ & $A_{\rho_V}^m$ & $A_{\rho_T}^m$ \\ \hline
%------------------------------------------------------ 1 --------------------------------------------------------------
$\hat{H}^1$ & $\tilde{M}\hat{M'}$ & $-10\theta$ & $32\theta$ & $6d\theta$ & $32\theta$ & $22d\theta$ \\ \hline
%------------------------------------------------------ 2 --------------------------------------------------------------
$\hat{H}^2$ & cons. & $-14\theta$ & $-4\theta$ & $6d\theta$ & $-4\theta$ & $6d\theta$ \\ \hline
%------------------------------------------------------ 3 --------------------------------------------------------------
$\hat{H}^3$ & $1/\hat{M}\hat{M'}$ & $-8\theta$ & $16\theta$ & $4d\theta$ & $16\theta$ & $8d\theta$ \\ \hline
%------------------------------------------------------ 4 --------------------------------------------------------------
$\hat{H}^4$ & $1/\hat{M}^2\hat{M'}^2$ & $-4\theta$ & $-8\theta$ & $4d\theta$ & $-8\theta$ & $4d\theta$ \\ \hline
%------------------------------------------------------ 5 --------------------------------------------------------------
$\hat{H}^5$ & $\hat{P} \hat{P'}$ & $14\phi$ & $4\phi$ & $20\mathop{pp'}\theta - 20(p^2 + \mathop{p'}^2)\phi - 6d\phi + 20\mathop{pp'}\eta$ & $4\phi$ & $24\mathop{pp'}\theta - 24(p^2 + \mathop{p'}^2)\phi -6d\phi + 24\mathop{pp'}\eta$ \\ \hline
%------------------------------------------------------ 6 --------------------------------------------------------------
$\hat{H}^6$ & $\hat{P}\hat{P'}/\hat{M}\hat{M'}$ & $16\phi$ & $-32\phi$ & $-8d\phi$ & $-32\phi$
& $-16d\phi$ \\ \hline
%------------------------------------------------------ 7 --------------------------------------------------------------
$\hat{H}^7$ & $\hat{P}\hat{P'}/\hat{M}^2\hat{M'}^2$ & $12\phi$ & $24\phi$ & $8\mathop{pp'}\theta - 8(p^2 + \mathop{p'}^2)\phi -12d\phi + 8\mathop{pp'}\eta$ & $24\phi$ & $16\mathop{pp'}\theta - 16(p^2 + \mathop{p'}^2)\phi - 12d\phi + 16\mathop{pp'}\eta$ \\ \hline
%------------------------------------------------------ 8 --------------------------------------------------------------
$\hat{H}^8$ & $\hat{P}^2\hat{P'}^2/\hat{M}\hat{M'}$ & $-8\eta$ & $16\eta$ & $4d\eta$ & $16\eta$ & $8d\eta$ \\ \hline
%------------------------------------------------------ 9 --------------------------------------------------------------
$\hat{H}^9$ & $\hat{P}^2\hat{P'}^2/\hat{M}^2\hat{M'}^2$ & $-12\eta$ & $-24\eta$ & $-16\mathop{pp'}\phi + 16(p^2 + \mathop{p'}^2)\eta + 12d\eta - 16\mathop{pp'}\varphi$ & $-24\eta$ & $-32\mathop{pp'}\phi + 32(p^2 + \mathop{p'}^2)\eta + 12d\eta - 32\mathop{pp'}\varphi$ \\ \hline
%----------------------------------------------------- 10 --------------------------------------------------------------
$\hat{H}^{10}$ & $\hat{P}^3\hat{P'}^3/\hat{M}^2\hat{M'}^2$ & $4\varphi$ & $8\varphi$ & $8\mathop{pp'}\eta - 8(p^2 + \mathop{p'}^2)\varphi - 4d\varphi + 8\mathop{pp'}\vartheta$ & $8\varphi$ & $16\mathop{pp'}\eta - 16(p^2 + \mathop{p'}^2)\varphi - 4d\varphi + 16\mathop{pp'}\vartheta$ \\ \hline
%----------------------------------------------------- 11 --------------------------------------------------------------
$\hat{H}^{11}$ & $\hat{P}^2\hat{M'}/\hat{M}$ & 0 & 0 & $4p^2\theta - 8\mathop{pp'}\phi + 4\mathop{p'}^2\eta$ & 0 & 0 \\ \hline
%----------------------------------------------------- 12 --------------------------------------------------------------
$\hat{H}^{12}$ & $\hat{P'}^2\hat{M}/\hat{M'}$ & 0 & 0 & $4\mathop{p'}^2\theta - 8\mathop{pp'}\phi +4p^2\eta$ & 0 & 0 \\
\end{tabular*}
      {\rule{\temptablewidth}{0.5pt}}
\end{center}
\end{table*}
%========================================================================================================================
\begin{table*}
\tabcolsep 1pt
\caption{Functions $\hat{J}^m$ and $B_{\lambda}^m$ ($m = 1, 2 ...12$) of $\lambda = \rho_{VT}$ couplings for calculating $\langle V_{E_\Delta}\rangle$.}
\vspace*{-12pt}
\begin{center}
\def\temptablewidth{0.96\textwidth}
{\rule{\temptablewidth}{0.5pt}}
\begin{tabular*}{\temptablewidth}{@{\extracolsep{\fill}}|c|c|c|c|c|c|c|c|}
%------------------------------------------------------ 1 --------------------------------------------------------------
$J^1$ & $\hat{P}\hat{M'}$ & $J^{7}$ & $\hat{P}^2\hat{P'}/\hat{M}$ & $B^1$ & $-44p\theta + 44\mathop{p'}\phi$ & $B^7$ & $8p\phi - 8\mathop{p'}\eta$ \\ \hline
%------------------------------------------------------ 2 --------------------------------------------------------------
$J^2$ & $\hat{P'}\hat{M}$ & $J^{8}$ & $\hat{P}\hat{P'}^2/\hat{M'}$ & $B^2$ & $-44\mathop{p'}\theta + 44p\phi$ & $B^8$ & $8\mathop{p'}\phi - 8p\eta$ \\ \hline
%------------------------------------------------------ 3 --------------------------------------------------------------
$J^3$ & $\hat{P}/\hat{M}$ & $J^{9}$ & $\hat{P}^2\hat{P'} /\hat{M}^2\hat{M'}$ & $B^3$ & $-8p\theta + 8\mathop{p'}\phi$ & $B^9$ & $48p\phi - 48\mathop{p'}\eta$ \\ \hline
%------------------------------------------------------ 4 --------------------------------------------------------------
$J^4$ & $\hat{P'}/\hat{M'}$ & $J^{10}$ & $\hat{P}\hat{P'}^2 /\hat{M}\hat{M'}^2$ & $B^4$ & $-8\mathop{p'}\theta + 8p\phi$ & $B^{10}$ & $48\mathop{p'}\phi - 48p\eta$ \\ \hline
%------------------------------------------------------ 5 --------------------------------------------------------------
$J^5$ & $\hat{P}/\hat{M}^2\hat{M'}$ & $J^{11}$ & $\hat{P}^3\hat{P'}^2 /\hat{M}^2\hat{M'}$ & $B^5$ & $-24p\theta + 24\mathop{p'}\phi$ & $B^{11}$ & $-24p\eta + 24\mathop{p'}\varphi$ \\ \hline
%------------------------------------------------------ 6 --------------------------------------------------------------
$J^6$ & $\hat{P'}/\hat{M}\hat{M'}^2$ & $J^{12}$ & $\hat{P}^2\hat{P'}^3 /\hat{M}\hat{M'}^2$ & $B^6$ & $-24\mathop{p'}\theta + 24p\phi$ & $B^{12}$ & $-24\mathop{p'}\eta + 24p\varphi$ \\
\end{tabular*}
      {\rule{\temptablewidth}{0.5pt}}
\end{center}
\end{table*}

Then the exchange term of energy for $\Delta$-isobars could be given as (here $i, j = \Delta^-, \Delta^0, \Delta^+, \Delta^{++}$):
\beq
\langle V_{E_\Delta}\rangle = \frac{1}{(2\pi)^4}\sum_\lambda\sum_{i,j} \int_0^{p_{F,i}}pdp \int_0^{p_{F,j}}\mathop{p'dp'} \sum_m \frac{1}{2}\left(\sum_{\lambda\neq\rho_{VT}}\Omega_{ij}^\lambda\Lambda_\lambda\hat{H}^m A^m_\lambda + \Lambda_{\rho_{VT}}\Omega_{ij}^\rho\hat{J}^m B^m\right),
\eeq
where the matrix element $\Omega_{ij}^\lambda$ is related to the isospin part and will be evaluated in Appendix C. $\Lambda_\lambda$ are the coupling constants of different meson with $\Delta$-isobars:
\beq
\Lambda_\sigma=\frac{1}{9}g_{\sigma_\Delta}^2 &,&
\Lambda_\omega=\frac{1}{9}g_{\omega_\Delta}^2,\\
\Lambda_\pi=\frac{1}{9}\left(\frac{f_{\pi_\Delta}}{m_\pi}\right)^2 &,& \Lambda_{\rho_T}=\frac{1}{9}\left(\frac{f_{\rho_\Delta}}{2M_\Delta}\right)^2,\\
\Lambda_{\rho_V}=\frac{1}{9}g_{\rho_\Delta}^2 &,&
\Lambda_{\rho_{VT}}=\frac{1}{9}\left(\frac{f_{\rho_\Delta}g_{\rho_\Delta}}{2M_\Delta}\right).
\eeq
$\hat{H}^m, A^m_\lambda, \hat{J}^m$ and $B^m_\lambda$ are the functions related to momenta and meson masses, as listed in the Tables II and III. The functions $\eta$, $\varphi$, $\vartheta$ in those tables are defined as:
\beq
\eta_\lambda(\mathop{p,p'})&=&\frac{\mathop{p^2 + p'^2} + m_\lambda^2}{\mathop{2pp'}}\phi_\lambda(\mathop{p,p'}),\\
\varphi_\lambda(\mathop{p,p'})&=&\frac{\mathop{p^2 + p'^2} + m_\lambda^2}{\mathop{2pp'}}\eta_\lambda(\mathop{p,p'}) + \frac{2}{3},\\
\vartheta_\lambda(\mathop{p,p'})&=&\frac{\mathop{p^2 + p'^2} + m_\lambda^2}{\mathop{2pp'}}\varphi_\lambda(\mathop{p,p'}).
\eeq

\section{Appendix~C}

%========================================================================================================================
\begin{table}\label{tab3}
\begin{center}
\caption{Matrix elements $\Omega_{kl}$ of isovector mesons for the isospin part in $\langle V_{E_N}\rangle$.}
\begin{tabular}{|c|c|c|}\hline
\diagbox{$l$}{$k$} & n & p \\ \hline
n & 1 & 2 \\ \hline
p & 2 & 1 \\ \hline
\end{tabular}
\end{center}
\end{table}

For the isoscalar mesons ($\sigma$, $\omega$), the isospin part of the exchange potential requires that the initial and final isospin wave functions to be identity, the matrix element $\Omega_{kl}$ then corresponds that of a unit matrix.

While for isovector mesons ($\rho, \pi$), the isospin part in the exchange contributions of Eqs.~(58-59) could be written as:
\beq
\chi_2^\dagger \bm{\tau}_\Delta\chi_1 \chi_2^\dagger \bm{\tau}_\Delta\chi_1,
\eeq
where $\chi_i$ is the isospin wave functions of baryon.

The isospin matrices for nucleons are the usual Pauli matrix.
Then the matrix element $\Omega_{kl}$ for nucleons can be evaluated (shown in Table IV).

The isospin matrices for $\Delta$-isobar are given as:
\beq
\bm{\tau}_{\Delta1} = \begin{pmatrix}0 & \sqrt{3} & 0 & 0 \\
\sqrt{3} & 0 & 2 & 0 \\
0 & 2 & 0 & \sqrt{3} \\
0 & 0 & \sqrt{3} & 0
\end{pmatrix},\quad
\bm{\tau}_{\Delta2} = \begin{pmatrix}0 & -\sqrt{3}i & 0 & 0 \\
\sqrt{3}i & 0 & -2i & 0 \\
0 & 2i & 0 & -\sqrt{3}i \\
0 & 0 & \sqrt{3}i & 0
\end{pmatrix},\quad
\bm{\tau}_{\Delta3} = \begin{pmatrix}3 & 0 & 0 & 0 \\
0 & 1 & 0 & 0 \\
0 & 0 & -1 & 0 \\
0 & 0 & 0 & -3
\end{pmatrix}.
\eeq
Then the matrix element $\Omega_{ij}$ for $\Delta$-isobars can be evaluated (shown in Table V).

\section{Appendix~D}
The components of self-energy for nucleons are given as (here $k,l$ denote neutron or proton):
\beq
\Sigma_S^k(p) &=& -\left(\frac{g_{\sigma_N}}{m_\sigma}\right)^2\rho_{s_N} - \frac{g_{\sigma N}g_{\sigma\Delta}}{m_\sigma^2} \rho_{S_\Delta} \nonumber \\
&&+ \frac{1}{(4\pi)^2}\frac{1}{p}\int_0^{p_F^k} \mathop{p'dp'} \left[\sum_{\substack{\lambda=\sigma,\omega,\\ \rho,\pi}} B_\lambda(p,\mathop{p'})\hat{M}(\mathop{p'})
+ \frac{1}{2}\hat{P}(\mathop{p'})D(p,\mathop{p'}) \right] \nonumber \\
&& + \frac{2}{(4\pi)^2}\frac{1}{p}\int_0^{p_F^{l(\neq k)}} \mathop{p'dp'}  \left[\sum_{\lambda'=\rho,\pi} B_{\lambda'}(p,\mathop{p'})\hat{M}(\mathop{p'})
+ \frac{1}{2}\hat{P}(\mathop{p'})D(p,\mathop{p'}) \right],
\eeq
\beq
\Sigma_0^k(p) &=& \left(\frac{g_{\omega_N}}{m_\omega}\right)^2\rho_{B_N} + \frac{g_{\omega N}g_{\omega\Delta}}{m_\omega^2} \rho_{B_\Delta} - \left(\frac{g_{\rho_N}}{m_\rho}\right)^2\rho_{B_{N3}}\tau_k - \frac{g_{\rho N}g_{\rho\Delta}}{m_\rho^2} \rho_{B_{\Delta 3}}\tau_k \nonumber \\
&& + \frac{1}{(4\pi)^2}\frac{1}{p}\int_0^{p_F^k}\mathop{p'dp'}
\sum_{\substack{\lambda=\sigma,\omega,\\ \rho,\pi}} A_\lambda(p,\mathop{p'}) \nonumber \\
&& + \frac{2}{(4\pi)^2}\frac{1}{p}\int_0^{p_F^{l(\neq k)}}\mathop{p'dp'}
\sum_{\lambda'=\rho,\pi} A_\lambda'(p,\mathop{p'}),
\eeq
\beq
\Sigma_V^k(p) &=& \frac{1}{(4\pi)^2}\frac{1}{p}\int_0^{p_F^k} \mathop{p'dp'} \left[\hat{P}(\mathop{p'})\sum_{\substack{\lambda=\sigma,\omega,\\ \rho,\pi}} C_\lambda(p,\mathop{p'})
+ \frac{1}{2}\hat{M}(\mathop{p'})D(p,\mathop{p'}) \right] \nonumber \\
&& + \frac{2}{(4\pi)^2}\frac{1}{p}\int_0^{p_F^{l(\neq k)}} \mathop{p'dp'}  \left[\hat{P}(\mathop{p'})\sum_{\lambda'=\rho,\pi} C_{\lambda'}(p,\mathop{p'})
+ \frac{1}{2}\hat{M}(\mathop{p'})D(p,\mathop{p'}) \right].
\eeq

For $\Delta$-isosbars, the three components are given as (here $i, j = \Delta^-, \Delta^0, \Delta^+, \Delta^{++}$ and $\lambda=\sigma,\omega,\rho,\pi$):
\beq
\Sigma_V^i(p) &= & \frac{1}{32\pi^2}\sum_j \Omega_{ij}\frac{1}{p}\int_0^{p_F^j}p'dp'\biggl[\sum_\lambda\Lambda_\lambda\biggl( \hat{P'}A_\lambda^5 + \frac{\hat{P'}}{\hat{M}\hat{M'}}A_\lambda^6 + \frac{\hat{P'}}{\hat{M}^2\hat{M'}^2}A_\lambda^7  + \frac{2\hat{P}\hat{P'}^2}{\hat{M}\hat{M'}}A_\lambda^8 + \frac{2\hat{P}\hat{P'}^2}{\hat{M}^2\hat{M'}^2}A_\lambda^9  + \frac{3\hat{P}^2\hat{P'}^3}{\hat{M}^2\hat{M'}^2}A_\lambda^{10} + \frac{2\hat{P}\hat{M'}}{\hat{M}}A_\lambda^{11}\biggr) \nonumber \\
&&+ \frac{1}{9}\frac{f_{\rho_\Delta}g_{\rho_\Delta}}{2M_\Delta}\biggl(\hat{M'}B^1 + \frac{1}{\hat{M}}B^3 + \frac{1}{\hat{M}^2\hat{M'}}B^5 + \frac{2\hat{P}\hat{P'}}{\hat{M}}B^7 + \frac{\hat{P'}^2}{\hat{M'}}B^8 + \frac{2\hat{P}\hat{P'}}{\hat{M}^2\hat{M'}}B^9 + \frac{\hat{P'}^2}{\hat{M}\hat{M'}^2}B^{10} + \frac{3\hat{P}^2\hat{P'}^2}{\hat{M}^2\hat{M'}}B^{11} + \frac{2\hat{P}\hat{P'}^3}{\hat{M}\hat{M'}^2}B^{12} \biggr)\biggr], \nonumber \\
\eeq
\beq
&&\Sigma_S^i(p) = - \frac{g_{\sigma_N}g_{\sigma_\Delta}}{m_\sigma^2}\rho_{S_N} - \frac{g_{\sigma_\Delta}^2}{m_\sigma^2}\rho_{S_\Delta} + \frac{1}{32\pi^2}\sum_j \Omega_{ij}\frac{1}{p}\int_0^{p_F^j}p'dp' \nonumber \\
&&\biggl[\sum_\lambda\Lambda_\lambda\biggl( \hat{M'}A_\lambda^1 - \frac{1}{\hat{M}^2\hat{M'}}A_\lambda^3 - \frac{2}{\hat{M}^3\hat{M'}^2}A_\lambda^4 - \frac{\hat{P}\hat{P'}}{\hat{M}^2\hat{M'}}A_\lambda^6 - \frac{2\hat{P}\hat{P'}}{\hat{M}^3\hat{M'}^2}A_\lambda^7 - \frac{\hat{P}^2\hat{P'}^2}{\hat{M}^2\hat{M'}}A_\lambda^8 - \frac{2\hat{P}^2\hat{P'}^2}{\hat{M}^3\hat{M'}^2}A_\lambda^9 - \frac{2\hat{P}^3\hat{P'}^3}{\hat{M}^3\hat{M'}^2}A_\lambda^{10} - \frac{\hat{P}^2\hat{M'}}{\hat{M}^2}A_\lambda^{11} + \frac{\hat{P'}^2}{\hat{M'}}A_\lambda^{12}\biggr) \nonumber \\
&& + \frac{1}{9}\frac{f_{\rho_\Delta}g_{\rho_\Delta}}{2M_\Delta}\biggl( \hat{P'}B^2 - \frac{\hat{P}}{\hat{M}^2}B^3
- \frac{2\hat{P}}{\hat{M}^3\hat{M'}}B^5 - \frac{\hat{P'}}{\hat{M}^2\hat{M'}^2}B^6 - \frac{\hat{P}^2\hat{P'}}{\hat{M}^2}B^7 - \frac{2\hat{P}^2\hat{P'}}{\hat{M}^3\hat{M'}}B^9 - \frac{\hat{P}\hat{P'}^2}{\hat{M}^2\hat{M'}^2}B^{10}  - \frac{\hat{P}^3\hat{P'}^2}{\hat{M}^3\hat{M'}}B^{11} - \frac{\hat{P}^2\hat{P'}^3}{\hat{M}^2\hat{M'}^2}B^{12}
\biggr)\biggr], \nonumber \\
\eeq
%-----------------------------------------------------------------------------------------------------------------------
\beq
\Sigma_0^i(p) &=& \frac{g_{\omega_N}g_{\omega_\Delta}}{m_\omega^2}\rho_{B_N} + \frac{g_{\rho_N}g_{\rho_\Delta}}{m_\rho^2}\rho_{B_{N3}}\tau_i + \frac{g_{\omega_\Delta}^2}{m_\omega^2}\rho_{B_\Delta} + \frac{g_{\rho_\Delta}^2}{m_\rho^2}\rho_{B_{\Delta3}}\tau_i + \frac{1}{32\pi^2}\sum_j \Omega_{ij}\frac{1}{p}\int_0^{p_F^j}p'dp' \nonumber \\
&&\biggl[\sum_\lambda\Lambda_\lambda(A_\lambda^2+\hat{H}^3 A_\lambda^3 + 3\hat{H}^4 A_\lambda^4 + \hat{H}^6 A_\lambda^6 + 2\hat{H}^7 A_\lambda^7 + \hat{H}^9 A_\lambda^9)+ \Lambda_{\rho_{TV}}(\hat{J}^3 B^3 + \hat{J}^4 B^4 + 2\hat{J}^5 B^5 + 2\hat{J}^6 B^6 + \hat{J}^9 B^9 + \hat{J}^{10} B^{10})\biggr]. \nonumber \\
\eeq

%========================================================================================================================
\begin{table}\label{tab1}
\begin{center}
\caption{Matrix elements $\Omega_{ij}$ of isovector mesons for the isospin part in $\langle V_{E_\Delta}\rangle$.}
\begin{tabular}{|c|c|c|c|c|}\hline
\diagbox{$j$}{$i$} & $\Delta^{++}$ & $\Delta^{+}$ & $\Delta^{0}$ & $\Delta^{-}$ \\ \hline
$\Delta^{++}$ & 9 & 6 & 0 & 0 \\ \hline
$\Delta^{+}$ & 6 & 1 & 8 & 0 \\ \hline
$\Delta^{0}$ & 0 & 8 & 1 & 6 \\ \hline
$\Delta^{-}$ & 0 & 0 & 6 & 9 \\ \hline
\end{tabular}
\end{center}
\end{table}
%------------------------------------------------------

\end{document}